# Generalized Multiple Access Channels with Confidential Messages [1] [2]

Yingbin Liang and H. Vincent Poor [3]

**Abstract**

A discrete memoryless generalized multiple access channel (GMAC) with confidential messages is studied, where two users attempt to transmit common information to a destination and each user also has private (confidential) information intended for the destination. This channel generalizes the multiple access channel (MAC) in that the two users also receive channel outputs, and hence may obtain the confidential information sent by each other from channel outputs they receive. However, each user views the other user as a wire-tapper, and wishes to keep its confidential information as secret as possible from the other user. The level of secrecy of the confidential information is measured by the equivocation rate, i.e., the entropy rate of the confidential information conditioned on channel outputs at the wire-tapper. The performance measure of interest for the GMAC with confidential messages is the rate-equivocation tuple that includes the common rate, two private rates and two equivocation rates as components. The set that includes all these achievable rate-equivocation tuples is referred to as the capacity-equivocation region. For the GMAC with one confidential message set, where only one user (user 1) has private (confidential) information for the destination, inner and outer bounds on the capacity-equivocation region are derived. These bounds match partially, and hence the capacity-equivocation region is partially characterized. Furthermore, the outer bound provides a tight converse to the secrecy capacity region, which is the set of all achievable rates with user 2 being perfectly ignorant of confidential messages of user 1, thus establishing the secrecy capacity region. A class of degraded GMACs with one confidential message set is further studied, and the capacity-equivocation region and the secrecy capacity region are established. These capacity results are further explored via two example degraded channels: the binary GMAC and the Gaussian GMAC. For both channels, the capacity-equivocation regions are obtained. In particular, the capacity-equivocation region of the degraded Gaussian GMAC is shown to apply to non-physically-degraded Gaussian channels as well. For the GMAC with two confidential message sets, where both users have confidential messages for the destination, an inner bound on the capacity-equivocation region is obtained. The secrecy rate region is derived, where each user's confidential information is perfectly hidden from the other user. It is demonstrated that for the case of two confidential message sets there is a trade-off between the two equivocation rates corresponding to the two confidential message sets, and this trade-off can be achieved by using codebooks that achieve different boundary points of the corresponding MAC.

---

[1] The material in this paper has been submitted in part to the IEEE International Symposium on Information Theory, Seattle, Washington, July 2006.

[2] This research was supported by the National Science Foundation under Grant ANI-03-38807.

[3] The authors are with the Department of Electrical Engineering, Princeton University, Engineering Quadrangle, Olden Street, Princeton, NJ 08544; e-mail: {`yingbinl,poor`}`@princeton.edu`

# 1 Introduction

Two important issues in communications are reliability and security. The reliability quantifies the maximum achievable rate (capacity) with small probability of error, and has been studied intensively since Shannon theory was established [1]. Security is an important issue when the transmitted information is confidential and needs to be kept as secret as possible from wire-tappers or eavesdroppers. The level of secrecy of confidential information at a wire-tapper can be measured by the equivocation rate, i.e., the entropy rate of confidential messages conditioned on channel outputs at the wire-tapper. If both reliability and security are considered, the performance measure of interest is the rate-equivocation tuple that includes both the communication rates and the equivocation rates (achieved at wire-tappers) as components. We refer to the set that consists of all achievable rate-equivocation tuples as the *capacity-equivocation region*.

Communication of confidential messages has been studied in the literature for some classes of channels. The wire-tap channel was introduced by Wyner in [2], where a sender wishes to transmit information to a legitimate receiver and to keep a wire-tapper as ignorant of this information as possible. The channel from the sender to the legitimate receiver and the wire-tapper was assumed to be a degraded broadcast channel. The trade-off between the communication rate to the legitimate receiver and the level of ignorance at the wire-tapper was developed. Furthermore, the secrecy capacity was established, at which the information source can be reliably reconstructed at the legitimate receiver with the wire-tapper being perfectly ignorant of the information source.

The broadcast channel with confidential messages was studied in [3] as a generalization of the wire-tap channel, where the sender also wishes to transmit common information to both the legitimate receiver and the wire-tapper in addition to the private (confidential) information to the legitimate receiver. Moreover, the broadcast channel from the sender to the two receivers was assumed to be general and may not be degraded. The capacity-equivocation region was established for this channel, and the secrecy capacity region was given. The relay channel with confidential messages was studied in [4], where the relay node acts as both a helper and a wire-tapper. Some other related studies on communication of confidential messages can be found in [5, 6, 7, 8, 9, 10, 11, 12, 13, 14].

In this paper, we consider a two-user *generalized multiple access channel (GMAC) with confidential messages*, which generalizes the multiple access channel (MAC) [15, Sec. 14.3] by allowing both users to receive noisy channel outputs. This channel model is motivated by wireless communications, where transmitted signals are broadcast over open media and can be received by all nodes within communication range. For this channel, we assume that two users (users 1 and 2) have common information and each user has its private (confidential) information intended for a destination. Since two users also receive channel outputs, they may extract each other's confidential information from their received channel outputs. However, each user treats the other user as a wire-tapper, and wishes to keep this wire-tapper as ignorant of its confidential message as possible. The level of ignorance of one user's confidential message at the other user (wire-tapper) is measured by the equivocation rate. Our goal is to study the capacity-equivocation region of the GMAC with confidential messages.

We first study the GMAC with one confidential message set, where two users have common information for the destination and only one user (user 1) has private (confidential) information for the destination. This channel generalizes the MAC with degraded message sets [16] with the further assumption that user 2 receives channel outputs and user 1 wants to keep user 2 as ignorant of its confidential information as possible. This model is also a counterpart of the broadcast channel with confidential messages studied in [3]. For the GMAC with one confidential message set, we obtain inner and outer bounds on the capacity-equivocation region. The two bounds match partially and determine the capacity-equivocation region partially. Furthermore, the outer bound provides a tight converse to the secrecy capacity region, which is the set of all achievable rates with user 2 being perfectly ignorant of confidential messages of user 1, and we hence establish the secrecy capacity region.

We further study the degraded GMAC with one confidential message set, where outputs at user 2 are degraded versions of outputs at the destination. This model generalizes the wire-tap channel [2] to allow user 1 and user 2 (the wire-tapper) to send common information to the destination. For the degraded GMAC with one confidential message set, we show a tight converse and establish the capacity-equivocation region and the secrecy capacity region. Moreover, we study these capacity results via two classes of degraded channels: the binary GMAC and the Gaussian GMAC. For both channels, we characterize the capacity-equivocation regions and secrecy capacity regions explicitly. In particular, for the Gaussian GMAC, we show that the capacity-equivocation region also applies to non-physically-degraded Gaussian channels.

We finally study the general case of the GMAC with two confidential message sets, where both users have confidential messages for the destination in addition to common messages. We obtain an achievable rate-equivocation region (inner bound on the capacity-equivocation region). We demonstrate a trade-off between the two equivocation rates corresponding to the two set of confidential messages sent by user 1 and user 2, and this trade-off can be achieved by using codebooks that achieve different boundary points of the corresponding MAC. This trade-off is a new feature that arises in the case with two confidential message sets. Based on the rate-equivocation region, we derive the secrecy rate region, where confidential messages of each user is perfectly secret from the other user.

In this paper, we adopt the following notation. We use upper case letters to indicate random variables, and we use lower case letters to indicate deterministic variables or realizations of the corresponding random variables. Exceptions will be clarified where they appear in the paper. We use $x^n$ to indicate the vector $(x_1,\ldots,x_n)$, and use $x_i^n$ to indicate the vector $(x_i,\ldots,x_n)$. Throughout the paper, the logarithmic function is to the base 2.

The organization of this paper is as follows. In Section 2, we introduce the channel model of the GMAC with confidential messages. In Section 3, we present our results for the GMAC with one confidential message set. In Section 4, we present our results for the degraded GMAC with one confidential message set and illustrate our results by a binary example channel. In Section 5, we focus on the Gaussian GMAC with one confidential message set. In Section 6, we present our results for the general case of the GMAC with two confidential message sets, and illustrate the intuition behind the result. In the final section, we give concluding remarks.

## 2  Channel Model

In this section, we first define the GMAC, and then define the performance measure of interest for the GMAC with confidential messages.

**Definition 1.** *A discrete memoryless GMAC consists of two finite channel input alphabets $\mathcal{X}_1$ and $\mathcal{X}_2$, three finite channel output alphabets $\mathcal{Y}, \mathcal{Y}_1$ and $\mathcal{Y}_2$, and a transition probability distribution $p(y, y_1, y_2 | x_1, x_2)$ (see Fig. 1), where $x_1 \in \mathcal{X}_1$ and $x_2 \in \mathcal{X}_2$ are channel inputs from users 1 and 2, respectively, and $y \in \mathcal{Y}, y_1 \in \mathcal{Y}_1$ and $y_2 \in \mathcal{Y}_2$ are channel outputs at the destination, user 1 and user 2, respectively.*

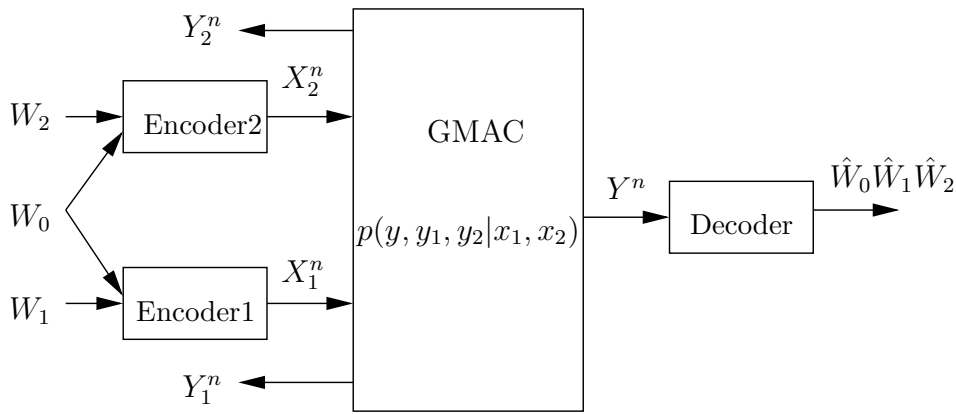

Figure 1: Generalized multiple access channel

For generality, we assume that each user receives channel outputs that also depend on its own inputs. This assumption is also practical since transmitted signals from one user may cause interference at its own receiver. All results that we obtain apply to the case where outputs at either user do not depend on its own inputs.

**Definition 2.** A $(2^{nR_0}, 2^{nR_1}, 2^{nR_2}, n)$ code for the GMAC consists of the following:

- Three message sets: $\mathcal{W}_0 = \{1, 2, \ldots, 2^{nR_0}\}$, $\mathcal{W}_1 = \{1, 2, \ldots, 2^{nR_1}\}$ and $\mathcal{W}_2 = \{1, 2, \ldots, 2^{nR_2}\}$. The common message $W_0$ and private messages $W_1$ and $W_2$ are independent and uniformly distributed over the message sets $\mathcal{W}_0, \mathcal{W}_1$ and $\mathcal{W}_2$, respectively.

- Two (stochastic) encoders, one at user 1: $\mathcal{W}_0 \times \mathcal{W}_1 \to \mathcal{X}_1^n$, which maps each message pair $(w_0, w_1) \in \mathcal{W}_0 \times \mathcal{W}_1$ to a codeword $x_1^n \in \mathcal{X}_1^n$; the other at user 2: $\mathcal{W}_0 \times \mathcal{W}_2 \to \mathcal{X}_2^n$, which maps each message pair $(w_0, w_2) \in \mathcal{W}_0 \times \mathcal{W}_2$ to a codeword $x_2^n \in \mathcal{X}_2^n$;

- One decoder at the destination: $\mathcal{Y}^n \to \mathcal{W}_0 \times \mathcal{W}_1 \times \mathcal{W}_2$, which maps a received sequence $y^n$ to a message tuple $(w_0, w_1, w_2) \in \mathcal{W}_0 \times \mathcal{W}_1 \times \mathcal{W}_2$.

Note that in the GMAC although users 1 and 2 can receive channel outputs (see Fig. 1), they are only passive listeners in that their encoding functions are not affected by these received outputs. However, since outputs at each user contain the other user's private (confidential) information, each user may extract the other user's private (confidential) information from its outputs. We assume that each user treats the other user as a wire-tapper, and wishes to keep the other user as ignorant of its private (confidential) messages as possible. We hence define the following two equivocation rates:

$$\text{at user 2: } \frac{1}{n}H(W_1|Y_2^n, X_2^n, W_0, W_2) \qquad (1)$$
$$\text{at user 1: } \frac{1}{n}H(W_2|Y_1^n, X_1^n, W_0, W_1)$$

which indicate the level of ignorance of the confidential message $W_1$ at user 2 and the level of ignorance of the confidential message $W_2$ at user 1, respectively. Note that the larger the equivocation rate, the higher the level of secrecy.

For the GMAC with confidential messages, a rate-equivocation tuple $(R_0, R_1, R_2, R_{1,e}, R_{2,e})$ is *achievable* if there exists a sequence of $(2^{nR_0}, 2^{nR_1}, 2^{nR_2}, n)$ codes with the average error probability $P_e^{(n)} \to 0$ as $n$ goes to infinity and with the equivocation rates $R_{1,e}$ and $R_{2,e}$ satisfying

$$\lim_{n\to\infty} \frac{1}{n}H(W_1|Y_2^n, X_2^n, W_0, W_2) \geq R_{1,e} \qquad (2)$$
$$\lim_{n\to\infty} \frac{1}{n}H(W_2|Y_1^n, X_1^n, W_0, W_1) \geq R_{2,e}.$$

Note that the rate-equivocation tuple $(R_0, R_1, R_2, R_{1,e}, R_{2,e})$ includes both the reliable communication rates and the equivocation rates, and it indicates the common and private rates $(R_0, R_1, R_2)$ achieved at certain levels of communication secrecy $(R_{1,e}, R_{2,e})$.

The capacity-equivocation region, denoted by $\mathscr{C}$, is the closure of the set that consists of all achievable rate-equivocation tuples $(R_0, R_1, R_2, R_{1,e}, R_{2,e})$. Our goal is to study the capacity-equivocation region of the GMAC with confidential messages.

# 3 GMAC with One Confidential Message Set

In this section, we study the GMAC with one confidential message set, where user 2 has only common messages and does not have confidential messages for the destination. This model generalizes the MAC with degraded message sets studied in [16] to consider the private messages sent from user 1 to be confidential, i.e., needing to be as secret as possible from user 2. This model is also a counterpart of the broadcast channel with confidential messages studied in [3].

In the following, we first provide inner and outer bounds on the capacity-equivocation region. We then present the secrecy capacity region, which includes all rates at which perfect secrecy can be achieved for private (confidential) messages sent by user 1. We finally give a proof of the outer bound on the capacity-equivocation region.

## 3.1 Main Results

For the GMAC with one confidential message set, the rate $R_2 = 0$, and the equivocation rate $R_{2,e}$ is not of interest. Hence channel outputs at user 1 do not play roles in the analysis. For notational convenience, we use $R_e$ to indicate $R_{1,e}$ in this case. Now the rate-equivocation tuple becomes $(R_0, R_1, R_e)$; i.e., it contains three components. We use $\mathscr{C}^I$ to denote the capacity-equivocation region of the GMAC in this situation.

The following two theorems provide inner and outer bounds on the capacity-equivocation region.

**Theorem 1.** *The following convexified region is an inner bound on the capacity-equivocation region for the GMAC with one confidential message set:*

$$\mathscr{R}^I = \textbf{Convex} \bigcup_{\substack{p(q,x_2)p(u|q)p(x_1|u)\\p(y,y_2|x_1,x_2)}} \left\{\begin{array}{l} (R_0, R_1, R_e): \\ R_0 \geq 0, \ R_1 \geq 0, \\ R_1 \leq I(U;Y|X_2,Q), \\ R_0 + R_1 \leq I(U,X_2,Q;Y), \\ 0 \leq R_e \leq R_1, \\ R_e \leq [I(U;Y|X_2,Q) - I(U;Y_2|X_2,Q)]_+, \\ R_e \leq [I(U,X_2,Q;Y) - R_0 - I(U;Y_2|X_2,Q)]_+ \end{array}\right\}. \tag{3}$$

*where the function $[x]_+ = x$ if $x \geq 0$ and $[x]_+ = 0$ if $x < 0$. The auxiliary random variables $Q$ and $U$ are bounded in cardinality by $|\mathcal{Q}| \leq |\mathcal{X}_1|\cdot|\mathcal{X}_2|+3$ and $|\mathcal{U}| \leq |\mathcal{X}_1|^2\cdot|\mathcal{X}_2|^2+4|\mathcal{X}_1|\cdot|\mathcal{X}_2|+3$, respectively.*

The proof of Theorem 1 follows from the achievable rate-equivocation region for the GMAC with two confidential message sets given in Theorem 7 in Section 6. This can be seen by setting $R_2 = 0$, $R_{2,e} = 0$, and $V := X_2$, and combining (75) with (80).

**Remark 1.** *The last bound in (3) indicates that there is a trade-off between the common rate and the secrecy level of confidential messages. As common rate $R_0$ increases, the secrecy level of confidential messages may get lower.*

**Remark 2.** *The rate-equivocation region (3) reduces to the capacity region of the MAC with degraded message sets given in [16] by setting $U = X_1$ and $R_e = 0$.*

**Theorem 2.** *The following region is an outer bound on the capacity-equivocation region of the GMAC with one confidential message set:*

$$\overline{\mathscr{R}}^I = \bigcup_{\substack{p(q,x_2)p(u|q)p(x_1|u) \\ p(v|q)p(y,y_2|x_1,x_2)}} \left\{ \begin{array}{l} (R_0, R_1, R_e): \\ R_0 \geq 0, \ R_1 \geq 0, \\ R_1 \leq I(U;Y|X_2,V), \\ R_0 + R_1 \leq I(U,X_2,Q;Y), \\ 0 \leq R_e \leq R_1, \\ R_e \leq I(U;Y|X_2,Q) - I(U;Y_2|X_2,Q), \\ R_0 + R_e \leq I(U,X_2,Q;Y) - I(U;Y_2|X_2,Q) \end{array} \right\}. \quad (4)$$

The proof of Theorem 2 is relegated to Section 3.3. In the following, we focus on the properties that the inner and outer bounds imply.

**Remark 3.** *The last four bounds in the outer bound (4) match the last four bounds in the inner bound (3), and hence these four common bounds partially determine the boundary of the capacity-equivocation region.*

We now study the case where perfect secrecy is achieved, i.e., user 2 does not get any information about confidential messages that user 1 sends to the destination. This happens if $R_e = R_1$.

**Definition 3.** *The* secrecy capacity region $\mathcal{C}_s^I$ *is the region that includes all achievable rate pairs $(R_0, R_1)$ such that $R_e = R_1$, i.e.,*

$$\mathcal{C}_s^I = \{(R_0, R_1) : (R_0, R_1, R_1) \in \mathscr{C}^I\}. \quad (5)$$

**Definition 4.** *For a given rate $R_0$, the* secrecy capacity *is the maximum achievable rate $R_1$ with confidential messages perfectly hidden from user 2, i.e.,*

$$C_s^I(R_0) = \max_{(R_0,R_1) \in \mathcal{C}_s^I} R_1. \quad (6)$$

Although the outer bound given in Theorem 2 provides only a partial converse to the capacity-equivocation region, it is sufficiently tight to serve as the converse to the secrecy capacity region and secrecy capacity (as a function of the common rate $R_0$).

**Theorem 3.** *For the GMAC with one confidential message set, the following region is the secrecy capacity region :*

$$\mathcal{C}_s^I = \bigcup_{\substack{p(q,x_2)p(u|q)p(x_1|u) \\ p(y,y_2|x_1,x_2)}} \left\{ \begin{array}{l} (R_0, R_1): \\ R_1 \leq I(U;Y|X_2,Q) - I(U;Y_2|X_2,Q), \\ R_0 + R_1 \leq I(U,X_2,Q;Y) - I(U;Y_2|X_2,Q) \end{array} \right\}. \quad (7)$$

*The secrecy capacity for a given rate $R_0$ is given by*

$$C_s^I(R_0) = \max \min \{ I(U;Y|X_2,Q) - I(U;Y_2|X_2,Q), \\ I(U,X_2,Q;Y) - I(U;Y_2|X_2,Q) - R_0 \} \quad (8)$$

*where the maximum is taken over all joint distributions $p(q,x_2)p(u|q)p(x_1|u)p(y,y_2|x_1,x_2)$. In both (7) and (8), the auxiliary random variables $Q$ and $U$ are bounded in cardinality by $|\mathcal{Q}| \leq |\mathcal{X}_1| \cdot |\mathcal{X}_2| + 3$ and $|\mathcal{U}| \leq |\mathcal{X}_1|^2 \cdot |\mathcal{X}_2|^2 + 4|\mathcal{X}_1| \cdot |\mathcal{X}_2| + 3$, respectively.*

*Proof.* The achievability of $\mathcal{C}_s^I$ follows from the inner bound on the capacity-equivocation region given in Theorem 1, and the converse follows from the last two bounds in the outer bound given in Theorem 2. The secrecy capacity $C_s^I(R_0)$ then easily follows from the secrecy capacity region $\mathcal{C}_s^I$. □

**Remark 4.** *If we let $R_0 = 0$ and $X_2 := \phi$, the GMAC with one confidential message set reduces to the case of a broadcast channel with confidential messages studied in [3] with the common rate being zero. For this channel, the secrecy capacity in (8) reduces to*

$$\mathcal{C}_s^I = \max[I(U;Y) - I(U;Y_2)] \quad (9)$$

*where the max is taken over all joint distribution $p(u,x_1)p(y,y_2|x_1)$. This is the same as the secrecy capacity given in Corollary 2 in [3].*

## 3.2 An Example

In this section, we consider an example of a discrete memoryless GMAC with one confidential message set. We obtain the capacity-equivocation region and the secrecy capacity region for this channel.

Consider a binary channel with all channel inputs and outputs having alphabets $\{0,1\}$. The MAC from two users to the destination is a binary multiplier channel, and the channel from user 1 to user 2 is a bias channel. The channel input-output relationship (see Fig. 2) satisfies

$$Y = X_1 \cdot X_2, \qquad Y_2 = \begin{cases} 1, & \text{if } X_1 \leq X_2; \\ 0, & \text{if } X_1 > X_2. \end{cases} \quad (10)$$

| $X_1$ | $X_2$ | $Y$ | $Y_2$ |
|---|---|---|---|
| 0 | 0 | 0 | 1 |
| 0 | 1 | 0 | 1 |
| 1 | 0 | 0 | 0 |
| 1 | 1 | 1 | 1 |

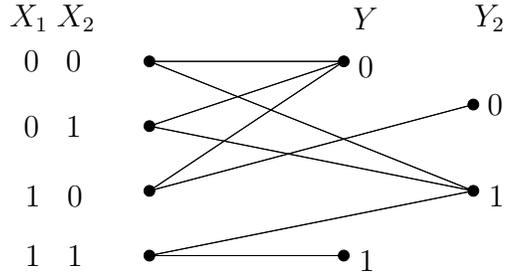

Figure 2: An example GMAC

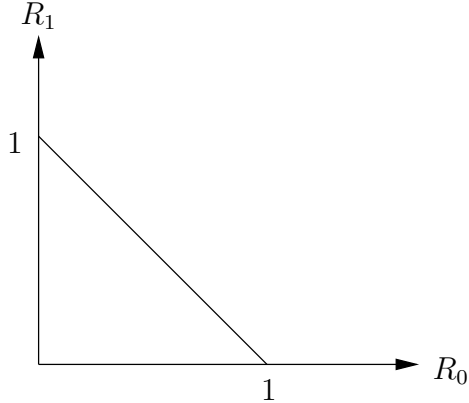

Figure 3: Capacity region of binary multiplier MAC

The capacity-equivocation region of the example channel given in (10) is:

$$\{(R_0, R_1, R_e) : R_0 + R_1 \leq 1, R_e = R_1\}. \tag{11}$$

The capacity-equivocation region implies that the secrecy capacity region of this channel is:

$$\{(R_0, R_1) : R_0 + R_1 \leq 1\}. \tag{12}$$

It is shown in [17] that the region $\{(R_0, R_1) : R_0 + R_1 \leq 1\}$ (see Fig. 3) is the capacity region of the binary multiplier MAC. We now show that perfect secrecy can be achieved for the two corner points of this region. It is trivial that perfect secrecy can be achieved for the corner point $(R_0 = 1, R_1 = 0)$, i.e., $R_e = 0$ is achievable at this point. For the other corner point $(R_0 = 0, R_1 = 1)$, perfect secrecy is achieved by sending $(x_1 = 0, x_2 = 1)$ for $W_1 = 0$ and $(x_1 = 1, x_2 = 1)$ for $W_1 = 1$. When either of these two codewords is transmitted, user 2 always gets output $Y_2 = 1$, and hence cannot determine whether $W_1 = 0$ or $W_1 = 1$ is sent. Therefore, perfect secrecy is achieved. By time-sharing between these two corner points, perfect secrecy can be achieved for the entire region, which is the best rate-equivocation region that can be achieved. Hence we obtain the capacity-equivocation region (11).

**Remark 5.** *The example channel given in (10) is a nondegraded channel. We hence obtain the capacity-equivocation region for a nondegraded channel.*

## 3.3 Proof of Theorem 2

In this section, we give a proof of the outer bound on the capacity-equivocation region in Theorem 2. The proof applies the techniques in the proofs of the converse of the capacity-equivocation region of the broadcast channel with confidential messages in [3] and the converse of the capacity region of the MAC [15, Chapter 14].

We consider a sequence of $(2^{nR_0}, 2^{nR_1}, n)$ codes for a GMAC with one confidential message set with $P_e^{(n)} \to 0$. Then the probability distribution on $W_0 \times W_1 \times \mathcal{X}_1^n \times \mathcal{X}_2^n \times \mathcal{Y}^n \times \mathcal{Y}_2^n$ is given by

$$p(w_0, w_1, x_1^n, x_2^n, y^n, y_2^n)$$
$$= p(w_0)p(w_1)p(x_1^n|w_0, w_1)p(x_2^n|w_0) \prod_{i=1}^{n} p(y_i, y_{2,i}|x_{1,i}, x_{2,i}). \tag{13}$$

By Fano's Inequality, we have

$$H(W_0, W_1|Y^n) \leq n(R_0 + R_1)P_e^{(n)} + 1 := n\delta_n \tag{14}$$

where $\delta_n \to 0$ if $P_e^{(n)} \to 0$.

We first give a lemma that is useful in the following proof.

**Lemma 1.** *[3, Lemma 7]*

$$\sum_{i=1}^{n} I(Z_i; Y^{i-1}|Z_{i+1}^n, T) = \sum_{i=1}^{n} I(Y_i; Z_{i+1}^n|Y^{i-1}, T).$$

We define the following auxiliary random variables.

$$Q_i := (Y^{i-1}, Y_{2,i+1}^n, X_2^n, W_0), \quad U_i := (W_1, Q_i), \quad V_i := (Y^{i-1}, X_2^n, W_0) \tag{15}$$

Note that these auxiliary random variables satisfy the following Markov chain conditions:

$$X_{2,i} \to Q_i \to U_i \to X_{1,i},$$
$$V_i \to Q_i \to (U_i, X_{1,i}, X_{2,i}) \tag{16}$$
$$(V_i, Q_i, U_i) \to (X_{1,i}, X_{2,i}) \to (Y_i, Y_{2,i})$$

We first consider

$$
\begin{aligned}
nR_{1,e} &\leq H(W_1|Y_2^n, X_2^n, W_0) \\
&= H(W_1|X_2^n, W_0) - I(W_1; Y_2^n|X_2^n, W_0) \\
&= I(W_1; Y^n|X_2^n, W_0) - I(W_1; Y_2^n|X_2^n, W_0) + H(W_1|Y^n, X_2^n, W_0) \\
&\overset{(a)}{\leq} \sum_{i=1}^{n} I(W_1; Y_i|Y^{i-1}, X_2^n, W_0) - I(W_1; Y_{2,i}|Y_{2,i+1}^n, X_2^n, W_0) + n\delta_n \\
&= \sum_{i=1}^{n} I(W_1, Y_{2,i+1}^n; Y_i|Y^{i-1}, X_2^n, W_0) - I(Y_{2,i+1}^n; Y_i|Y^{i-1}, X_2^n, W_0, W_1) \\
&\quad - I(W_1, Y^{i-1}; Y_{2,i}|Y_{2,i+1}^n, X_2^n, W_0) + I(Y^{i-1}; Y_{2,i}|Y_{2,i+1}^n, X_2^n, W_0, W_1) + n\delta_n \\
&\overset{(b)}{=} \sum_{i=1}^{n} I(W_1, Y_{2,i+1}^n; Y_i|Y^{i-1}, X_2^n, W_0) - I(W_1, Y^{i-1}; Y_{2,i}|Y_{2,i+1}^n, X_2^n, W_0) + n\delta_n \\
&= \sum_{i=1}^{n} I(Y_{2,i+1}^n; Y_i|Y^{i-1}, X_2^n, W_0) + I(W_1; Y_i|Y^{i-1}, Y_{2,i+1}^n, X_2^n, W_0) \\
&\quad - I(Y^{i-1}; Y_{2,i}|Y_{2,i+1}^n, X_2^n, W_0) - I(W_1; Y_{2,i}|Y^{i-1}, Y_{2,i+1}^n, X_2^n, W_0) + n\delta_n \\
&\overset{(c)}{=} \sum_{i=1}^{n} I(W_1; Y_i|Y^{i-1}, Y_{2,i+1}^n, X_2^n, W_0) - I(W_1; Y_{2,i}|Y^{i-1}, Y_{2,i+1}^n, X_2^n, W_0) + n\delta_n \\
&\overset{(d)}{=} \sum_{i=1}^{n} I(U_i; Y_i|X_{2,i}, Q_i) - I(U_i; Y_{2,i}|X_{2,i}, Q_i) + n\delta_n
\end{aligned}
\tag{17}
$$

In the preceding equation, $(a)$ follows from the chain rule and Fano's inequality (14), $(b)$ and $(c)$ follows from Lemma 1, and $(d)$ follows from the definition for $Q_i$ and $U_i$ in (15).

We also can write

$$
\begin{aligned}
nR_0 &+ nR_{1,e} \\
&\leq H(W_0) + H(W_1|Y_2^n, X_2^n, W_0) \\
&\leq I(W_0; Y^n) + H(W_1|Y_2^n, X_2^n, W_0) + n\delta_n \\
&= \sum_{i=1}^{n} I(W_0; Y_i|Y^{i-1}) + H(W_1|Y_2^n, X_2^n, W_0) + n\delta_n \\
&\overset{(a)}{\leq} \sum_{i=1}^{n} I(Y^{i-1}, Y_{2,i+1}^n, X_2^n, W_0; Y_i) + H(W_1|Y_2^n, X_2^n, W_0) + n\delta_n \\
&\overset{(b)}{\leq} \sum_{i=1}^{n} I(Q_i, X_{2,i}; Y_i) + I(U_i; Y_i|X_{2,i}, Q_i) - I(U_i; Y_{2,i}|X_{2,i}, Q_i) + n\delta_n \\
&= \sum_{i=1}^{n} I(X_{2,i}, Q_i, U_i; Y_i) - I(U_i; Y_{2,i}|X_{2,i}, Q_i) + n\delta_n
\end{aligned}
\tag{18}
$$

In the preceding equation, $(a)$ follows from the chain rule and nonnegativity of mutual information, and $(b)$ follows from the definition for $Q_i$ in (15) and the bound (17).

We further have

$$\begin{aligned}
nR_1 = H(W_1) &\leq I(W_1; Y^n) + n\delta_n \leq I(W_1; Y^n|W_0) + n\delta_n \\
&= H(W_1|W_0) - H(W_1|Y^n, W_0) + n\delta_n \\
&\overset{(a)}{\leq} H(W_1|W_0, X_2^n) - H(W_1|Y^n, W_0, X_2^n) + n\delta_n \\
&= I(W_1; Y^n|W_0, X_2^n) + n\delta_n \\
&= \sum_{i=1}^n I(W_1; Y_i|Y^{i-1}, W_0, X_2^n) + n\delta_n \\
&\leq \sum_{i=1}^n I(Y^{i-1}, Y_{2,i+1}^n, X_2^n, W_0, W_1; Y_i|Y^{i-1}, X_2^n, W_0) + n\delta_n \\
&= \sum_{i=1}^n I(U_i; Y_i|X_{2,i}, V_i) + n\delta_n
\end{aligned} \quad (19)$$

where $(a)$ follows from the fact that $W_1$ is independent of $(W_0, X_2^n)$ and conditioning does not increase entropy.

Finally, we have

$$\begin{aligned}
nR_0 + nR_1 = H(W_0, W_1) &\leq I(W_0, W_1; Y^n) + n\delta_n \\
&= \sum_{i=1}^n I(W_0, W_1; Y_i|Y^{i-1}) + n\delta_n \\
&\overset{(a)}{\leq} \sum_{i=1}^n I(Y^{i-1}, Y_{2,i+1}^n, X_2^n, W_0, W_1; Y_i) + n\delta_n \\
&= \sum_{i=1}^n I(U_i, Q_i, X_{2,i}; Y_i) + n\delta_n
\end{aligned} \quad (20)$$

where $(a)$ follows from the chain rule and nonnegativity of mutual information. Theorem 2 then follows from standard single letter characterization (see e.g. [15]).

## 4 Degraded GMAC with One Confidential Message Set

In this section, we study the degraded GMAC with one confidential message set, where the output at user 2 is a degraded version of the output at the destination. This channel generalizes the wire-tap channel studied in [2] to allow user 1 and user 2 (wire-tapper) to jointly send common information to the destination. In the following, we first present the main results that characterize the capacity-equivocation region and the secrecy capacity region of the degraded channel, and provide the proof for the main results. We then illustrate our results via a binary example GMAC.

## 4.1 Main Results

We first define two classes of degraded GMACs with one confidential message set in the following.

**Definition 5.** *The GMAC with one confidential message set is physically degraded if the transition probability distribution satisfies*

$$p(y, y_2|x_1, x_2) = p(y|x_1, x_2)p(y_2|y, x_2), \tag{21}$$

*i.e., $y_2$ is independent of $x_1$ conditioned on $y$ and $x_2$.*

**Definition 6.** *The GMAC with one confidential message set is stochastically degraded if its conditional marginal distribution is the same as that of a physically degraded GMAC, i.e., there exists a distribution $p(y_2|y, x_2)$ such that*

$$p(y_2|x_1, x_2) = \sum_y p(y|x_1, x_2)p(y_2|y, x_2). \tag{22}$$

We note the following useful lemma.

**Lemma 2.** *The capacity-equivocation region of GMACs with confidential messages depends only on the marginal channel transition probability distributions $p(y|x_1, x_2)$, $p(y_1|x_1, x_2)$, and $p(y_2|x_1, x_2)$.*

*Proof.* It is clear that the decoding probability of error at the destination depends only on the probability distribution $p(y|x_1, x_2)$, and so is the achievable rates $(R_0, R_1, R_2)$. Moreover, the equivocation rates $R_{1,e}$ and $R_{2,e}$ depend only on the probability distributions $p(y_1|x_1, x_2)$ and $p(y_2|x_1, x_2)$, respectively. □

Based on Lemma 2, we have the following capacity-equivocation region for both physically and stochastically degraded GMACs with confidential messages.

**Theorem 4.** *For the degraded GMAC with one confidential message set, the capacity-equivocation region is given by*

$$\mathscr{C}_d^I = \bigcup_{\substack{p(q, x_2)p(x_1|q) \\ p(y|x_1, x_2)p(y_2|y, x_2)}} \left\{ \begin{array}{l} (R_0, R_1, R_e): \\ R_0 \geq 0, R_1 \geq 0, \\ R_1 \leq I(X_1; Y|X_2, Q), \\ R_0 + R_1 \leq I(X_1, X_2; Y), \\ 0 \leq R_e \leq R_1, \\ R_e \leq I(X_1; Y|X_2, Q) - I(X_1; Y_2|X_2, Q), \\ R_0 + R_e \leq I(X_1, X_2; Y) - I(X_1; Y_2|X_2, Q) \end{array} \right\}. \tag{23}$$

*where $Q$ is bounded in cardinality by $|\mathcal{Q}| \leq |\mathcal{X}_1| \cdot |\mathcal{X}_2| + 1$.*

The achievability is obtained by applying Theorem 1 and setting $U = X_1$ in (3). The proof of the converse is provided in the next subsection.

**Remark 6.** *The region $\mathscr{C}_d^I$ can be shown to be convex (similar to the proof of Lemma 5 in [3]), and hence does not need further convexification.*

The following results on the secrecy capacity region and secrecy capacity (as a function of $R_0$) follow from Theorem 4.

**Corollary 1.** *For the degraded GMAC with one confidential message set, the secrecy capacity region is given by*

$$\mathcal{C}_s = \bigcup_{\substack{p(q, x_2) p(x_1|q) \\ p(y|x_1, x_2) p(y_2|y, x_2)}} \left\{ \begin{array}{l} (R_0, R_1): \\ R_1 \leq I(X_1; Y|X_2, Q) - I(X_1; Y_2|X_2, Q), \\ R_0 + R_1 \leq I(X_1, X_2; Y) - I(X_1; Y_2|X_2, Q) \end{array} \right\}. \quad (24)$$

*The secrecy capacity as a function of $R_0$ is given by*

$$C_s(R_0) = \max \min\{ I(X_1; Y|X_2, Q) - I(X_1; Y_2|X_2, Q), \\ I(X_1, X_2; Y) - I(X_1; Y_2|X_2, Q) - R_0 \} \quad (25)$$

*where the maximum is taken over all joint distributions $p(q, x_2)p(x_1|q)p(y|x_1, x_2)p(y_2|y, x_2)$. In both (24) and (25), $Q$ is bounded in cardinality by $|\mathcal{Q}| \leq |\mathcal{X}_1| \cdot |\mathcal{X}_2| + 1$.*

**Remark 7.** *If $R_0 = 0$ and $X_2 := \phi$, the degraded GMAC with one confidential message set reduces to the wire-tap channel studied in [2]. For this channel, the secrecy capacity in (24) reduces to*

$$\mathcal{C}_s = \max[I(X_1; Y) - I(X_1; Y_2)] \quad (26)$$

*where the max is taken over all joint distribution $p(x_1)p(y|x_1)p(y_2|y)$. This is consistent with the secrecy capacity given in [2] with a different form, because the problem in [2] is formulated as a source-channel problem. The secrecy capacity in (26) is also consistent with the secrecy capacity of the less noisy channel given in [3, Theorem 3], because degraded channels belong to the class of less noisy channels [18].*

## 4.2 Proof of the Converse for Theorem 4

For the general discrete memoryless GMAC with one confidential message set, we give a proof of the outer bound on the capacity-equivocation region in Section 3.3. This outer bound provides only a partial converse. In this section, we apply the degradedness condition and prove a tight converse to the capacity-equivocation region for the degraded GMAC with one confidential message set.

Our proof applies the techniques in the converse proofs for the wire-tap channel in [2] and for the MAC in [15, Chapter 14].

We consider a sequence of $(2^{nR_0}, 2^{nR_1}, n)$ codes for the GMAC with one confidential message set with $P_e^{(n)} \to 0$. Then the probability distribution on $W_0 \times W_1 \times \mathcal{X}_1^n \times \mathcal{X}_2^n \times \mathcal{Y}^n \times \mathcal{Y}_2^n$ is given by

$$p(w_0, w_1, x_1^n, x_2^n, y^n, y_2^n)$$
$$= p(w_0)p(w_1)p(x_1^n|w_0, w_1)p(x_2^n|w_0) \prod_{i=1}^n p(y_i, y_{2,i}|x_{1,i}, x_{2,i}) \quad (27)$$
$$= p(w_0)p(w_1)p(x_1^n|w_0, w_1)p(x_2^n|w_0) \prod_{i=1}^n p(y_i|x_{1,i}, x_{2,i})p(y_{2,i}|y_i, x_{2,i}).$$

By Fano's inequality, we have

$$H(W_0, W_1|Y^n) \leq n(R_0 + R_1)P_e^{(n)} + 1 := n\delta_n \quad (28)$$

where $\delta_n \to 0$ if $P_e^{(n)} \to 0$.

we define the following auxiliary random variable:

$$Q_i := (Y^{i-1}, X_2^n, W_0). \quad (29)$$

Note that $X_{2,i} \to Q_i \to X_{1,i}$ and $Q_i \to (X_{1,i}, X_{2,i}) \to (Y_i, Y_{2,i})$ form Markov chains.

We first consider

$$nR_{1,e} = H(W_1|Y_2^n, X_2^n, W_0)$$
$$= H(W_1|Y_2^n, X_2^n, W_0) - H(W_1|Y_2^n, X_2^n, W_0, Y^n) + H(W_1|Y_2^n, X_2^n, W_0, Y^n)$$
$$\overset{(a)}{\leq} I(W_1; Y^n|Y_2^n, X_2^n, W_0) + n\delta_n$$
$$\leq I(W_1, X_1^n; Y^n|Y_2^n, X_2^n, W_0) + n\delta_n$$
$$= I(X_1^n; Y^n|Y_2^n, X_2^n, W_0) + I(W_1; Y^n|Y_2^n, X_1^n, X_2^n, W_0) + n\delta_n$$
$$\overset{(b)}{=} I(X_1^n; Y^n|Y_2^n, X_2^n, W_0) + n\delta_n$$
$$= H(X_1^n|Y_2^n, X_2^n, W_0) - H(X_1^n|Y^n, Y_2^n, X_2^n, W_0) + n\delta_n$$
$$\overset{(c)}{=} H(X_1^n|Y_2^n, X_2^n, W_0) - H(X_1^n|Y^n, X_2^n, W_0) + n\delta_n$$
$$= I(X_1^n; Y^n|X_2^n, W_0) - I(X_1^n; Y_2^n|X_2^n, W_0) + n\delta_n$$

where $(a)$ follows from Fano's inequality, $(b)$ follows from the fact that $Y^n$ is independent of $W_0, W_1$ given $Y_2^n, X_1^n, X_2^n$, and $(c)$ follows from the degradedness condition, i.e., $Y_2^n$ is independent of $X_1^n, W_0$ given $Y^n, X_2^n$.

We proceed to bound $nR_{1,e}$ and obtain

$$
\begin{aligned}
nR_{1,e} &\leq \sum_{i=1}^{n} I(X_1^n; Y_i | Y^{i-1}, X_2^n, W_0) - I(X_1^n; Y_{2,i} | Y_2^{i-1}, X_2^n, W_0) + n\delta_n \\
&= \sum_{i=1}^{n} H(Y_i | Y^{i-1}, X_2^n, W_0) - H(Y_i | Y^{i-1}, X_1^n, X_2^n, W_0) \\
&\quad - H(Y_{2,i} | Y_2^{i-1}, X_2^n, W_0) - H(Y_{2,i} | Y_2^{i-1}, X_1^n, X_2^n, W_0) + n\delta_n \\
&\stackrel{(d)}{\leq} \sum_{i=1}^{n} H(Y_i | Y^{i-1}, X_2^n, W_0) - H(Y_i | Y^{i-1}, X_{1,i}, X_2^n, W_0) \\
&\quad - H(Y_{2,i} | Y^{i-1}, Y_2^{i-1}, X_2^n, W_0) + H(Y_{2,i} | Y^{i-1}, X_{1,i}, X_2^n, W_0) + n\delta_n \\
&\stackrel{(e)}{=} \sum_{i=1}^{n} H(Y_i | Y^{i-1}, X_2^n, W_0) - H(Y_i | Y^{i-1}, X_{1,i}, X_2^n, W_0) \\
&\quad - H(Y_{2,i} | Y^{i-1}, X_2^n, W_0) + H(Y_{2,i} | Y^{i-1}, X_{1,i}, X_2^n, W_0) + n\delta_n \\
&\stackrel{(f)}{=} \sum_{i=1}^{n} H(Y_i | Q_i, X_{2,i}) - H(Y_i | Q_i, X_{2,i}, X_{1,i}) \\
&\quad - H(Y_{2,i} | Q_i, X_{2,i}) + H(Y_{2,i} | Q_i, X_{2,i}, X_{1,i}) + n\delta_n \\
&= \sum_{i=1}^{n} I(X_{1,i}; Y_i | Q_i, X_{2,i}) - I(X_{1,i}; Y_{2,i} | Q_i, X_{2,i}) + n\delta_n
\end{aligned}
\tag{30}
$$

In the preceding equation, the second term of $(d)$ follows from the fact that $Y_i$ is independent of everything else given $X_{1,i}$ and $X_{2,i}$; the third term of $(d)$ follows from the fact that conditioning does not increase entropy, and the fourth term of $(d)$ follows from the fact that $Y_{2,i}$ is independent of everything else given $X_{1,i}$ and $X_{2,i}$. In $(e)$, the third term follows from the degraded condition, i.e., $Y_2^{i-1}$ is independent of everything else given $Y^{i-1}, X_2^{i-1}$. The step $(f)$ follows from the definition of $Q_i$ given in (29).

$$
\begin{aligned}
nR_0 + nR_{1,e} &= H(W_0) + nR_{1,e} \stackrel{(a)}{\leq} I(W_0; Y^n) + nR_{1,e} + n\delta_n \\
&= \sum_{i=1}^{n} I(W_0; Y_i | Y^{i-1}) + nR_{1,e} + n\delta_n \\
&\stackrel{(b)}{\leq} \sum_{i=1}^{n} I(Y^{i-1}, X_2^n, W_0; Y_i) + nR_{1,e} + n\delta_n \\
&\stackrel{(c)}{\leq} \sum_{i=1}^{n} I(X_{2,i}, Q_i; Y_i) + I(X_{1,i}; Y_i | Q_i, X_{2,i}) - I(X_{1,i}; Y_{2,i} | Q_i, X_{2,i}) + n\delta_n \\
&\stackrel{(d)}{=} \sum_{i=1}^{n} I(X_{1,i}, X_{2,i}; Y_i) - I(X_{1,i}; Y_{2,i} | Q_i, X_{2,i}) + n\delta_n
\end{aligned}
\tag{31}
$$

where $(a)$ follows from Fano's inequality, $(b)$ follows from the chain rule and nonnegativity of mutual information, (c) follows from the inequality (30), and (d) used the Markov chain condition $Q_i \to (X_{1,i}, X_{2,i}) \to (Y_i, Y_{2,i})$.

The next two inequalities follow the converse proof for the MAC in [15]. For completeness, we include those steps here.

$$\begin{aligned}
nR_1 &\stackrel{(a)}{\leq} I(W_1; Y^n | W_0, X_2^n) + n\delta_n \\
&\leq I(X_1^n, W_1; Y^n | W_0, X_2^n) + n\delta_n \\
&= I(X_1^n; Y^n | W_0, X_2^n) + I(W_1; Y^n | W_0, X_2^n, X_1^n) + n\delta_n \\
&\stackrel{(b)}{\leq} I(X_1^n; Y^n | W_0, X_2^n) + n\delta_n \\
&= \sum_{i=1}^n I(X_1^n; Y_i | Y^{i-1}, W_0, X_2^n) + n\delta_n \\
&= \sum_{i=1}^n H(Y_i | Y^{i-1}, X_2^n, W_0) - H(Y_i | Y^{i-1}, X_2^n, W_0, X_1^n) + n\delta_n \\
&\leq \sum_{i=1}^n H(Y_i | Y^{i-1}, X_2^n, W_0) - H(Y_i | Y^{i-1}, X_2^n, W_0, X_{1,i}) + n\delta_n \\
&= \sum_{i=1}^n H(Y_i | X_{2,i}, Q_i) - H(Y_i | X_{2,i}, Q_i, X_{1,i}) + n\delta_n \\
&= \sum_{i=1}^n I(X_{1,i}; Y_i | X_{2,i}, Q_i) + n\delta_n
\end{aligned} \quad (32)$$

where (a) follows from the partial step in (19), and (b) follows from the Markov chain condition $(W_0, W_1) \to (X_{1,i}, X_{2,i}) \to Y_i$.

$$\begin{aligned}
nR_0 + nR_1 &\leq I(W_0, W_1; Y^n) + n\delta_n \\
&= \sum_{i=1}^n I(W_0, W_1; Y_i | Y^{i-1}) + n\delta_n \\
&\leq \sum_{i=1}^n H(Y_i | Y^{i-1}) - H(Y_i | Y^{i-1}, W_0, W_1, X_{2,i}, X_{1,i}) + n\delta_n \\
&\leq \sum_{i=1}^n H(Y_i) - H(Y_i | X_{2,i}, X_{1,i}) + n\delta_n \\
&= \sum_{i=1}^n I(X_{1,i}, X_{2,i}; Y_i) + n\delta_n.
\end{aligned} \quad (33)$$

The converse for Theorem 4 then follows by standard single letter characterization (see e.g. [15]).

## 4.3 A Binary GMAC with One Confidential Message Set

In this section, we study a binary GMAC with one confidential message set, which is a degraded channel. We first follow [19] to introduce notation and useful lemmas for binary channels. We then introduce the binary GMAC model we study and present the capacity-equivocation region for this channel.

We first define the following operation:

$$a * b := a(1-b) + (1-a)b \qquad \text{for } 0 \leq a, b \leq 1. \tag{34}$$

We then define the entropy function

$$h(a) := \begin{cases} -a \log a - (1-a) \log(1-a), & \text{if } 0 < a < 1; \\ 0, & \text{if } a = 0 \text{ or } 1. \end{cases} \tag{35}$$

The function $h(a)$ is one-to-one for $0 \leq a \leq 1/2$. The inverse of the entropy function is limited to $h^{-1}(c) \in [0, 1/2]$.

**Lemma 3.** *[19] The function $f(u) = h(\rho * h^{-1}(u)), 0 \leq u \leq 1$ (where $\rho \in (0, 1/2]$ is a fixed parameter) is strictly convex in $u$.*

The following useful lemma is a binary version of the entropy power inequality.

**Lemma 4.** *[19] Consider two binary random vectors $X^n$ and $Y^n$. Let $H(X^n) \geq nv$. Let*

$$Y_i = X_i \oplus Z_i \qquad \text{for } i = 1, \ldots, n \tag{36}$$

*where $Z^n$ is a binary random vector with i.i.d. components and $Z_i$ has distribution $Pr(Z_i = 1) = p_0$ where $0 < p_0 \leq 1/2$. The vectors $X^n$ and $Y^n$ can be viewed as inputs and outputs of a binary symmetric channel (BSC) with crossover probability $p_0$. Then,*

$$H(Y^n) \geq nh(p_0 * h^{-1}(v)) \tag{37}$$

*with equality if and only if $X^n$ has independent components, and $H(X_i) = v$ for $i = 1, 2, \ldots, n$.*

We now introduce the binary GMAC model of interest. We assume all channel inputs and outputs have the binary alphabet set $\{0, 1\}$. We assume that the channel is discrete memoryless and the input-output relationship at each time instant satisfies

$$Y_i = X_{1,i} \cdot X_{2,i}, \qquad Y_{2,i} = Y_i \oplus Z_{2,i} \qquad \text{for } i = 1, \ldots, n \tag{38}$$

where $Z_2^n$ is a binary random vector with i.i.d. components and $Z_{2,i}$ has distribution $Pr(Z_{2,i} = 1) = p$ where $0 < p \leq 1/2$. We illustrate the channel input-output relationship in Fig. 4. Note that the MAC channel from $(X_1, X_2)$ to $Y$ is a binary multiplier channel. It is clear that this GMAC channel is degraded, and the channel outputs $Y$ and $Y_2$ can be viewed as the input and output of a discrete memoryless BSC with crossover probability $p$.

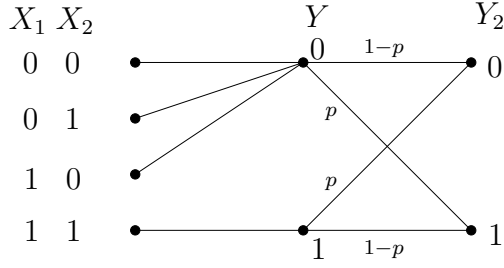

Figure 4: A degraded binary example GMAC

We have the following theorem on the capacity-equivocation region of the binary example GMAC with one confidential message set.

**Theorem 5.** *For the binary GMAC with one confidential message set given in* (38), *the capacity-equivocation region is given by*

$$\mathscr{C}^B = \bigcup_{0 \leq \alpha \leq \frac{1}{2}} \left\{ \begin{array}{l} (R_0, R_1, R_e): \\ R_0 \geq 0, R_1 \geq 0, \\ R_1 \leq h(\alpha), \\ R_0 + R_1 \leq 1, \\ 0 \leq R_e \leq R_1, \\ R_e \leq h(\alpha) + h(p) - h(p * \alpha), \\ R_0 + R_e \leq 1 + h(p) - h(p * \alpha) \end{array} \right\}. \tag{39}$$

**Corollary 2.** *The secrecy capacity region of the binary GMAC with one confidential message given in* (38) *is*

$$\mathcal{C}_s^B = \bigcup_{0 \leq \alpha \leq \frac{1}{2}} \left\{ \begin{array}{l} (R_0, R_1): \\ R_0 \geq 0, R_1 \geq 1, \\ R_1 \leq h(\alpha) + h(p) - h(p * \alpha), \\ R_0 + R_1 \leq 1 + h(p) - h(p * \alpha) \end{array} \right\}. \tag{40}$$

*The secrecy capacity as a function of $R_0$ is given by*

$$C_s^B(R_0) = h(\alpha^*) + h(p) - h(p * \alpha^*) \tag{41}$$

*where $\alpha^*$ is determined by the following equation*

$$R_0 = 1 - h(\alpha^*). \tag{42}$$

**Remark 8.** *The BSC crossover probability parameter $p$ determines how noisy the channel from user 1 to user 2 is compared to the channel from user 1 to the destination. When $p = 0$, user 2 has the same channel from user 1 as the destination, and hence no secrecy can be achieved. As $p$ increases, user 2 has a noisier channel from user 1 than the destination, and hence higher secrecy can be achieved. As $p = \frac{1}{2}$, user 2 is totally confused by confidential messages sent by user 1, and perfect secrecy is achieved.*

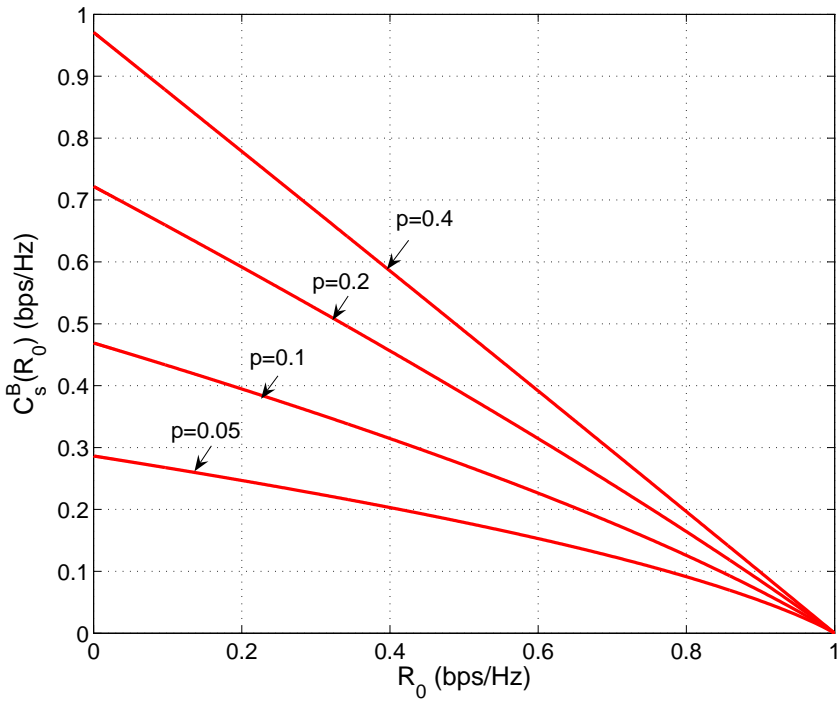

Figure 5: Secrecy capacity regions of the binary GMAC with one confidential message set

Fig. 5 plots the secrecy capacity as a function of $R_0$ for four values of $p$. These lines of $C_s^B(R_0)$ also serves as boundaries of the secrecy capacity regions for the binary channel we study with the vertical axis being viewed as $R_1$. It is clear from Fig. 5 that as $p$ increases, the secrecy capacity region enlarges, because user 2 is further confused by confidential messages sent by user 1.

**Remark 9.** *From the achievability proof of Theorem 5 (given in Section 4.4), it can be seen that the optimal scheme to achieve the secrecy capacity region uses superposition encoding. To achieve the secrecy capacity corresponding to different values of $R_0$, different values of the superposition parameter $\alpha$ needs to be chosen to generate the codebook. However, if the secrecy constraint is not considered, the capacity region of the binary multiplier MAC can be achieved by a time sharing scheme and superposition encoding is not necessary.*

Fig. 6 plots the secrecy capacity as a function of $R_0$ (indicated by the solid line) and compares it with the secrecy rate achieved by the time sharing scheme (indicated by the dashed line). The figure demonstrates that the time sharing scheme is strictly suboptimal to provide the secrecy capacity region. As we commented in Remark 9, although the time sharing scheme is optimal to achieve the capacity region of the binary multiplier MAC, it is not optimal to achieve the secrecy capacity region of the binary GMAC, as secrecy is considered as a performance criterion.

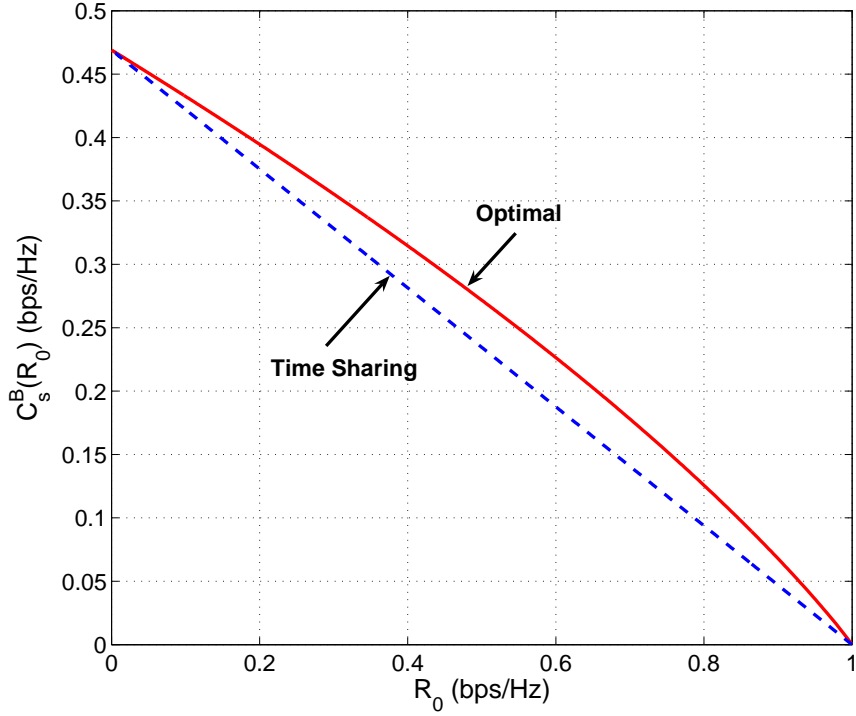

Figure 6: Comparison of secrecy capacity region and secrecy rate region achieved by time sharing scheme for the binary example GMAC with one confidential message set

## 4.4 Proof of Theorem 5

*Proof of the Achievability.* We apply Theorem 4 to prove that the capacity-equivocation region (39) is achievable. Let $Q$ and $X'$ be two binary random variables with alphabet $\{0, 1\}$, and assume that $Q$ is independent of $X'$. We choose the following joint distribution:

$$Pr\{Q = 0\} = \frac{1}{2}; \qquad Pr\{X' = 1\} = \alpha, \ 0 \leq \alpha \leq \frac{1}{2}; \qquad (43)$$
$$Pr\{X_2 = 1\} = 1; \qquad X_1 = Q \oplus X'.$$

We now compute the mutual information terms in the achievable region given in Theorem 4 based on the preceding joint distribution.

$$\begin{aligned}
R_1 &\leq I(X_1; Y | X_2, Q) = H(Y | X_2, Q) \\
&= Pr\{Q = 0\} H(Y | X_2 = 1, Q = 0) + Pr\{Q = 1\} H(Y | X_2 = 1, Q = 1) \\
&= h(\alpha) \\
R_0 + R_1 &\leq I(X_1, X_2; Y) = H(Y) = 1
\end{aligned}$$

$$\begin{aligned}
R_e &\leq I(X_1;Y|X_2,Q) - I(X_1;Y_2|X_2,Q) \\
&= h(\alpha) - (H(Y_2|X_2,Q) - H(Y_2|X_1,X_2)) \\
&= h(\alpha) - \big[Pr\{Q=0\}H(Y_2|X_2=1,Q=0) + Pr\{Q=1\}H(Y_2|X_2=1,Q=1) \\
&\quad - Pr\{X_1=0\}H(Y_2|X_2=1,X_1=0) - Pr\{X_1=1\}H(Y_2|X_2=1,X_1=1)\big] \\
&= h(\alpha) - [h(\alpha * p) - h(p)] \\
&= h(\alpha) + h(p) - h(\alpha * p) \\
R_0 + R_1 &\leq I(X_1,X_2;Y) - I(X_1;Y_2|X_2,Q) = 1 - [h(\alpha * p) - h(p)] \\
&= 1 + h(p) - h(\alpha * p)
\end{aligned}$$

□

*Proof of the Converse.* We apply the converse bounds obtained in Section 4.2, and further derive these bounds for the example binary GMAC.

From the first step in (32), we obtain

$$nR_1 \leq I(X_1^n; Y^n | X_2^n, W_0) + n\delta_n = H(Y^n | X_2^n, W_0) + n\delta_n \qquad (44)$$

where we have used the deterministic property of the GMAC, which implies $H(Y^n | X_1^n, X_2^n, W_0) = 0$.

Since $\{Y_i, 1 \leq i \leq n\}$ are binary random variables, $H(Y_i) \leq 1$ for $1 \leq i \leq n$. Hence

$$0 \leq H(Y^n | X_2^n, W_0) \leq \sum_{i=1}^n H(Y_i) \leq n. \qquad (45)$$

It is clear that there exists a parameter $\alpha \in [0, 1/2]$ such that

$$H(Y^n | X_2^n, W_0) = nh(\alpha). \qquad (46)$$

Substituting the preceding equation into (44), we obtain

$$nR_1 \leq nh(\alpha) + n\delta_n. \qquad (47)$$

From (33), we obtain

$$\begin{aligned}
nR_0 + nR_1 &\leq I(W_0, W_1; Y^n) + n\delta_n \leq H(Y^n) + n\delta_n \\
&\leq n + n\delta_n.
\end{aligned} \qquad (48)$$

From (30), we obtain

$$\begin{aligned}
nR_{1,e} &\leq I(X_1^n; Y^n | X_2^n, W_0) - I(X_1^n; Y_2^n | X_2^n, W_0) + n\delta_n \\
&= H(Y^n | X_2^n, W_0) - H(Y_2^n | X_2^n, W_0) + H(Y_2^n | X_1^n, X_2^n, W_0) + n\delta_n \\
&\stackrel{(a)}{=} nh(\alpha) - H(Y_2^n | X_2^n, W_0) + H(Y_2^n | Y^n, X_1^n, X_2^n, W_0) + n\delta_n \qquad (49) \\
&\stackrel{(b)}{=} nh(\alpha) - H(Y_2^n | X_2^n, W_0) + H(Y_2^n | Y^n) + n\delta_n \\
&= nh(\alpha) - H(Y_2^n | X_2^n, W_0) + nh(p) + n\delta_n
\end{aligned}$$

In the preceding bound, the first term in $(a)$ follows from (46), the third term in $(a)$ follows from the fact that $Y^n$ is a deterministic function of $(X_1^n, X_2^n)$, and the third term in $(b)$ follows from the fact that $Y_2^n$ is conditionally independent of everything else given $Y_n$.

Since $Z_2^n$ in (38) is independent of $W_0, X_2^n$ and $Y^n$, we apply Lemma 4 to bound the term $H(Y_2^n|X_2^n, W_0)$.

$$\begin{aligned}
H(Y_2^n|X_2^n, W_0) &= \mathrm{E} H(Y_2^n|X_2^n = x_2^n, W_0 = w_0) \\
&\stackrel{(a)}{\geq} \mathrm{E}\left[nh\left(p * h^{-1}\left(\frac{H(Y^n|X_2^n = x_2^n, W_0 = w_0)}{n}\right)\right)\right] \\
&\stackrel{(b)}{\geq} nh\left(p * h^{-1}\left(\mathrm{E}\frac{H(Y^n|X_2^n = x_2^n, W_0 = w_0)}{n}\right)\right) \\
&= nh\left(p * h^{-1}\left(\frac{H(Y^n|X_2^n, W_0)}{n}\right)\right) \\
&\stackrel{(c)}{=} nh\left(p * h^{-1}\left(\frac{nh(\alpha)}{n}\right)\right) \\
&= nh(p * \alpha)
\end{aligned} \tag{50}$$

where $(a)$ follows from Lemma 4, $(b)$ follows from Lemma 3 and Jensen's inequality, and $(c)$ follows from (46).

Substituting (50) into (49), we obtain

$$nR_{1,e} \leq nh(\alpha) + nh(p) - nh(p * \alpha) + n\delta_n \tag{51}$$

From (31), we obtain

$$\begin{aligned}
nR_0 + nR_{1,e} &\leq I(W_0; Y^n) + nR_{1,e} \\
&\stackrel{(a)}{\leq} I(W_0; Y^n) + I(X_1^n; Y^n|X_2^n, W_0) - I(X_1^n; Y_2^n|X_2^n, W_0) + n\delta_n \\
&\stackrel{(b)}{\leq} I(W_0, X_2^n; Y^n) + I(X_1^n; Y^n|X_2^n, W_0) - I(X_1^n; Y_2^n|X_2^n, W_0) + n\delta_n \\
&= I(W_0, X_1^n, X_2^n; Y^n) - I(X_1^n; Y_2^n|X_2^n, W_0) + n\delta_n \\
&= H(Y^n) - I(X_1^n; Y_2^n|X_2^n, W_0) + n\delta_n \\
&\stackrel{(c)}{\leq} n + nh(p) - nh(p * \alpha) + n\delta_n
\end{aligned} \tag{52}$$

where $(a)$ follows from (30), $(b)$ follows from the chain rule and nonnegativity of mutual information, and $(c)$ follows from partial results in deriving $R_{1,e}$. $\square$

## 5 Gaussian GMAC with One Confidential Message Set

In this section, we study the Gaussian GMAC with one confidential message set, where the channel outputs at the destination and user 2 are corrupted by additive Gaussian noise terms.

We assume that the channel is discrete and memoryless, and that the channel input-output relationship at each time instant is given by

$$\begin{aligned} Y_i &= X_{1,i} + X_{2,i} + Z_i \\ Y_{2,i} &= X_{1,i} + X_{2,i} + Z_{2,i} \end{aligned} \quad (53)$$

where $Z^n$ and $Z_2^n$ are independent zero mean Gaussian random vectors with i.i.d. components. We assume that $Z_i$ and $Z_{2,i}$ have variances $N$ and $N_2$, respectively, where $N < N_2$. The channel input sequences $X_1^n$ and $X_2^n$ are subject to the average power constraints $P_1$ and $P_2$, respectively, i.e.,

$$\frac{1}{n}\sum_{i=1}^{n} X_{1,i}^2 \leq P_1, \quad \text{and} \quad \frac{1}{n}\sum_{i=1}^{n} X_{2,i}^2 \leq P_2. \quad (54)$$

The following theorem states the capacity-equivocation region for the Gaussian GMAC with one confidential message set.

**Theorem 6.** *For the Gaussian GMAC with one confidential message set given in* (53), *the capacity-equivocation region is given by*

$$\mathscr{C}^G = \bigcup_{0 \leq \alpha \leq 1} \left\{ \begin{array}{l} (R_0, R_1, R_e): \\ R_0 \geq 0, R_1 \geq 0, \\ R_1 \leq \frac{1}{2}\log\left(1 + \frac{\alpha P_1}{N}\right), \\ R_0 + R_1 \leq \frac{1}{2}\log\left(1 + \frac{P_1 + P_2 + 2\sqrt{\bar{\alpha} P_1 P_2}}{N}\right), \\ 0 \leq R_e \leq R_1, \\ R_e \leq \frac{1}{2}\log\left(1 + \frac{\alpha P_1}{N}\right) - \frac{1}{2}\log\left(1 + \frac{\alpha P_1}{N_2}\right), \\ R_0 + R_e \leq \frac{1}{2}\log\left(1 + \frac{P_1 + P_2 + 2\sqrt{\bar{\alpha} P_1 P_2}}{N}\right) - \frac{1}{2}\log\left(1 + \frac{\alpha P_1}{N_2}\right) \end{array} \right\}. \quad (55)$$

where $\bar{\alpha} = 1 - \alpha$ indicating the correlation between the inputs from users 1 and 2.

**Corollary 3.** *The secrecy capacity region of the Gaussian GMAC with one confidential message set given in* (53) *is*

$$\mathcal{C}_s^G = \bigcup_{0 \leq \alpha \leq 1} \left\{ \begin{array}{l} (R_0, R_1): \\ R_0 \geq 0, R_1 \geq 0, \\ R_1 \leq \frac{1}{2}\log\left(1 + \frac{\alpha P_1}{N}\right) - \frac{1}{2}\log\left(1 + \frac{\alpha P_1}{N_2}\right), \\ R_0 + R_1 \leq \frac{1}{2}\log\left(1 + \frac{P_1 + P_2 + 2\sqrt{\bar{\alpha} P_1 P_2}}{N}\right) - \frac{1}{2}\log\left(1 + \frac{\alpha P_1}{N_2}\right) \end{array} \right\}. \quad (56)$$

*The secrecy capacity as a function of $R_0$ is*

$$C_s^G(R_0) = \begin{cases} \frac{1}{2}\log\left(1 + \frac{P_1}{N}\right) - \frac{1}{2}\log\left(1 + \frac{P_1}{N_2}\right), & \text{if } R_0 \leq \frac{1}{2}\log\frac{P_1 + P_2 + N}{P_1 + N} \\ \frac{1}{2}\log\left(1 + \frac{\alpha^* P_1}{N}\right) - \frac{1}{2}\log\left(1 + \frac{\alpha^* P_1}{N_2}\right) & \text{if } R_0 > \frac{1}{2}\log\frac{P_1 + P_2 + N}{P_1 + N} \end{cases} \quad (57)$$

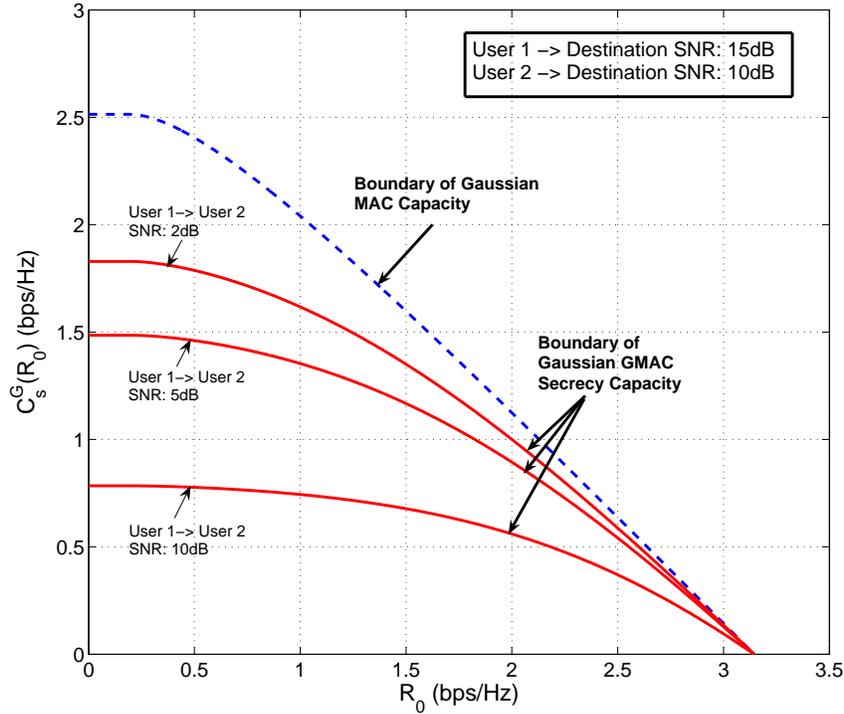

Figure 7: Secrecy capacity regions of Gaussian GMACs with one confidential message set and capacity region of corresponding Gaussian MAC

where $\alpha^*$ is determined by the following equation

$$R_0 = \frac{1}{2} \log \frac{P_1 + P_2 + 2\sqrt{(1-\alpha^*)P_1 P_2} + N}{\alpha^* P_1 + N}. \tag{58}$$

Fig. 7 plots the secrecy capacity $C_s^G(R_0)$ of Gaussian GMACs with one confidential message set for three user 1-to-user 2 SNR values. The lines of $C_s^G(R_0)$ also serve as boundaries of the secrecy capacity regions if we view the vertical axis as $R_1$. It can be seen that as user 1-to-user 2 SNR decreases, which implies that the noise level at user 2 increases, user 2 gets more confused by confidential messages sent by user 1. Thus the secrecy capacity region enlarges. As this SNR approaches zero, the secrecy capacity region approaches the entire capacity region of the Gaussian MAC, which means that perfect secrecy is achieved for almost all points in the capacity region of the MAC.

## 5.1 Proof of Theorem 6

To show Theorem 6, we first note that the Gaussian GMAC defined in (53) is not physically degraded according to Definition 5. However, it is stochastically degraded according to Definition 6, because the marginal distribution $p(y_2|x_1, x_2)$ is the same as that of the following physically degraded Gaussian GMAC:

$$\begin{aligned} Y_i &= X_{1,i} + X_{2,i} + Z_i \\ Y_{2,i} &= X_{1,i} + X_{2,i} + Z_i + Z'_i \end{aligned} \quad (59)$$

where $Z^n$ is the same as in (53). The random vector $Z'^n$ is independent of $Z^n$, and has i.i.d. components with each component having the distribution $\mathcal{N}(0, N_2 - N)$. According to Lemma 2, it is sufficient to prove Theorem 6 for the physically degraded Gaussian GMAC defined in (59).

*Proof of the Achievability.* The achievability follows by computing the mutual information terms in Theorem 4 with the following joint distribution:

$$\begin{aligned} Q &= \phi, \quad X_2 \sim \mathcal{N}(0, P_2) \\ X'_1 &\sim \mathcal{N}(0, \alpha P_1), \text{ and } X'_1 \text{ is independent of } X_2 \\ X_1 &= \sqrt{\frac{\bar{\alpha} P_1}{P_2}} X_2 + X'_1 \end{aligned} \quad (60)$$

□

*Proof of the Converse.* We apply the converse proof for the general degraded GMAC in Section 4.2, and further derive these bounds for the degraded Gaussian GMAC.

From (32), we obtain

$$\begin{aligned} nR_1 &\leq \sum_{i=1}^{n} I(X_{1,i}; Y_i | X_{2,i}, Q_i) + n\delta_n \\ &= \sum_{i=1}^{n} h(Y_i | X_{2,i}, Q_i) - h(Y_i | X_{1,i}, X_{2,i}, Q_i) + n\delta_n \\ &= \sum_{i=1}^{n} h(Y_i | X_{2,i}, Q_i) - h(Z_i | X_{1,i}, X_{2,i}, Q_i) + n\delta_n \\ &= \sum_{i=1}^{n} h(Y_i | X_{2,i}, Q_i) - \frac{1}{2} \log 2\pi e N + n\delta_n \end{aligned} \quad (61)$$

For the first term in the preceding inequality, we have

$$\begin{aligned} \sum_{i=1}^{n} h(Y_i | X_{2,i}, Q_i) &= \sum_{i=1}^{n} h(X_{1,i} + X_{2,i} + Z_i | X_{2,i}, Q_i) = \sum_{i=1}^{n} h(X_{1,i} + Z_i | X_{2,i}, Q_i) \\ &\leq \sum_{i=1}^{n} h(X_{1,i} + Z_i) \leq \sum_{i=1}^{n} \frac{1}{2} \log 2\pi e (\mathrm{E} X_{1,i}^2 + N) \\ &\stackrel{(a)}{\leq} \frac{n}{2} \log 2\pi e \left( \frac{1}{n} \sum_{i=1}^{n} \mathrm{E} X_{1,i}^2 + N \right) \leq \frac{n}{2} \log 2\pi e (P_1 + N) \end{aligned} \quad (62)$$

where $(a)$ follows from Jensen's inequality.

On the other hand,

$$\sum_{i=1}^n h(Y_i|X_{2,i}, Q_i) \geq \sum_{i=1}^n h(X_{1,i} + X_{2,i} + Z_i|X_{1,i}, X_{2,i}, Q_i) = \frac{n}{2} \log 2\pi eN . \tag{63}$$

Combining (62) and (63), we establish that there exists some $\alpha \in [0, 1]$ such that

$$\sum_{i=1}^n h(Y_i|X_{2,i}, Q_i) = \frac{n}{2} \log 2\pi e(\alpha P_1 + N) . \tag{64}$$

We hence obtain the bound for $R_1$

$$\begin{aligned} nR_1 &\leq \frac{n}{2} \log 2\pi e(\alpha P_1 + N) - \frac{1}{2} \log 2\pi eN + n\delta_n \\ &= \frac{n}{2} \log \left(1 + \frac{\alpha P_1}{N}\right) + n\delta_n . \end{aligned} \tag{65}$$

For the term $\sum_{i=1}^n h(Y_i|X_{2,i}, Q_i)$, we also derive the following bound:

$$\begin{aligned} &\sum_{i=1}^n h(Y_i|X_{2,i}, Q_i) \\ &= \sum_{i=1}^n h(X_{1,i} + Z_i|X_{2,i}, Q_i) \\ &\leq \sum_{i=1}^n \mathrm{E}_{X_{2,i},Q_i} \frac{1}{2} \log 2\pi e \mathsf{Var}(X_{1,i} + Z_i|X_{2,i}, Q_i) \\ &\stackrel{(a)}{\leq} \sum_{i=1}^n \frac{1}{2} \log 2\pi e \mathrm{E}_{X_{2,i},Q_i} \mathsf{Var}(X_{1,i} + Z_i|X_{2,i}, Q_i) \\ &= \sum_{i=1}^n \frac{1}{2} \log 2\pi e \left(\mathrm{E}_{X_{2,i},Q_i} \mathsf{Var}(X_{1,i}|X_{2,i}, Q_i) + N\right) \\ &= \sum_{i=1}^n \frac{1}{2} \log 2\pi e \left(\mathrm{E}(X_{1,i}^2) - \mathrm{E}_{X_{2,i},Q_i} \mathrm{E}^2(X_{1,i}|X_{2,i}, Q_i) + N\right) \\ &\stackrel{(b)}{\leq} \frac{n}{2} \log 2\pi e \left(\frac{1}{n}\sum_{i=1}^n \mathrm{E}(X_{1,i}^2) - \frac{1}{n}\mathrm{E}_{X_{2,i},Q_i} \mathrm{E}^2(X_{1,i}|X_{2,i}, Q_i) + N\right) \\ &\leq \frac{n}{2} \log 2\pi e \left(P_1 - \frac{1}{n}\mathrm{E}_{X_{2,i},Q_i} \mathrm{E}^2(X_{1,i}|X_{2,i}, Q_i) + N\right) \end{aligned} \tag{66}$$

where $(a)$ and $(b)$ follows from Jensen's inequality.

Using (64), we have

$$\alpha P_1 + N \leq P_1 - \frac{1}{n}\mathrm{E}_{X_{2,i},Q_i}\mathrm{E}^2(X_{1,i}|X_{2,i},Q_i) + N \tag{67}$$
$$\implies \frac{1}{n}\mathrm{E}_{X_{2,i},Q_i}\mathrm{E}^2(X_{1,i}|X_{2,i},Q_i) \leq \bar{\alpha}P_1$$

From (33), we obtain

$$\begin{aligned}
nR_0 + nR_1 &\leq \sum_{i=1}^{n} I(X_{1,i},X_{2,i};Y_i) + n\delta_n \\
&= \sum_{i=1}^{n} H(Y_i) - I(Y_i|X_{1,i},X_{2,i}) + n\delta_n \\
&= \sum_{i=1}^{n} H(Y_i) - \frac{n}{2}\log 2\pi eN + n\delta_n
\end{aligned} \tag{68}$$

For the first term in the preceding inequality, we obtain

$$\begin{aligned}
\sum_{i=1}^{n} h(Y_i) &= \sum_{i=1}^{n} h(X_i + X_{1,i} + Z_i) \\
&\leq \sum_{i=1}^{n} \frac{1}{2}\log 2\pi e\left(\mathrm{E}(X_{1,i}+X_{2,i})^2 + N\right) \\
&\stackrel{(a)}{\leq} \frac{n}{2}\log 2\pi e\left(\frac{1}{n}\sum_{i=1}^{n}\mathrm{E}(X_{1,i}+X_{2,i})^2 + N\right) \\
&\leq \frac{n}{2}\log 2\pi e\left(\frac{1}{n}\sum_{i=1}^{n}\mathrm{E}X_{1,i}^2 + \frac{1}{n}\sum_{i=1}^{n}\mathrm{E}X_{2,i}^2 + \frac{1}{n}\sum_{i=1}^{n}2\mathrm{E}(X_{1,i}X_{2,i}) + N\right) \\
&\leq \frac{n}{2}\log 2\pi e\left(P_1 + P_2 + \frac{1}{n}\sum_{i=1}^{n}2\mathrm{E}(X_{1,i}X_{2,i}) + N\right) \\
&\leq \frac{n}{2}\log 2\pi e\left(P_1 + P_2 + \frac{1}{n}\sum_{i=1}^{n}2\mathrm{E}_{X_{2,i},Q_i}\big(X_{2,i}\mathrm{E}(X_{1,i}|X_{2,i},Q_i)\big) + N\right) \\
&\stackrel{(b)}{\leq} \frac{n}{2}\log 2\pi e\left(P_1 + P_2 + \frac{2}{n}\sum_{i=1}^{n}\sqrt{\mathrm{E}_{X_{2,i},Q_i}X_{2,i}^2 \cdot \mathrm{E}_{X_{2,i},Q_i}\mathrm{E}^2(X_{1,i}|X_{2,i},Q_i)} + N\right) \\
&\stackrel{(c)}{\leq} \frac{n}{2}\log 2\pi e\bigg(P_1 + P_2 \\
&\qquad + 2\sqrt{\left(\frac{1}{n}\sum_{i=1}^{n}\mathrm{E}_{X_{2,i},Q_i}X_{2,i}^2\right)\cdot\left(\frac{1}{n}\sum_{i=1}^{n}\mathrm{E}_{X_{2,i},Q_i}\mathrm{E}^2(X_{1,i}|X_{2,i},Q_i)\right)} + N\bigg) \\
&\stackrel{(d)}{\leq} \frac{n}{2}\log 2\pi e\left(P_1 + P_2 + 2\sqrt{\bar{\alpha}P_1P_2} + N\right)
\end{aligned} \tag{69}$$

In the preceding bound, $(a)$ follows from Jensen's inequality, $(b)$ and $(c)$ follows from Cauchy-Schwarz inequality, and $(d)$ follows from (67).

Hence,
$$\begin{aligned} nR_0 + nR_1 &\leq \frac{n}{2}\log 2\pi e\left(P_1 + P_2 + 2\sqrt{\bar{\alpha}P_1 P_2} + N\right) - \frac{n}{2}\log 2\pi e N + n\delta_n \\ &= \frac{n}{2}\log\left(1 + \frac{P_1 + P_2 + 2\sqrt{\bar{\alpha}P_1 P_2}}{N}\right) + n\delta_n \end{aligned} \quad (70)$$

From (30), we obtain
$$\begin{aligned} nR_{1,e} &\leq \sum_{i=1}^{n} I(X_{1,i}; Y_i|X_{2,i}, Q_i) - I(X_{1,i}; Y_{2,i}|X_{2,i}, Q_i) + n\delta_n \\ &= \frac{n}{2}\log\left(1 + \frac{\alpha P_1}{N}\right) - \sum_{i=1}^{n} h(Y_{2,i}|X_{2,i}, Q_i) + \frac{n}{2}\log 2\pi e N_2 + n\delta_n \end{aligned} \quad (71)$$

To bound the term $\sum_{i=1}^{n} h(Y_{2,i}|X_{2,i}, Q_i)$ in (71), we first derive the following bound. Since $Z'_i$ is independent of $Y_i$ given $X_{2,i}$ and $Q_i$, by entropy power inequality, we obtain
$$\begin{aligned} 2^{2h(Y_i+Z'_i|X_{2,i}=x_{2,i},Q_i=q_i)} &\geq 2^{2h(Y_i|X_{2,i}=x_{2,i},Q_i=q_i)} + 2^{2h(Z'_i|X_{2,i}=x_{2,i},Q_i=q_i)} \\ &= 2^{2h(Y_i|X_{2,i}=x_{2,i},Q_i=q_i)} + 2\pi e(N_2 - N) \end{aligned}$$

We then obtain
$$h(Y_i + Z'_i|X_{2,i} = x_{2,i}, Q_i = q_i) \geq \frac{1}{2}\log\left(2^{2h(Y_i|X_{2,i}=x_{2,i},Q_i=q_i)} + 2\pi e(N_2 - N)\right)$$

Taking the expectation on both sides of the preceding equation, we obtain
$$\begin{aligned} \mathrm{E}h(Y_i + Z'_i|X_{2,i} = x_{2,i}, Q_i = q_i) &\geq \frac{1}{2}\mathrm{E}\log\left(2^{2h(Y_i|X_{2,i}=x_{2,i},Q_i=q_i)} + 2\pi e(N_2 - N)\right) \\ &\stackrel{(a)}{\geq} \frac{1}{2}\log\left(2^{2\mathrm{E}h(Y_i|X_{2,i}=x_{2,i},Q_i=q_i)} + 2\pi e(N_2 - N)\right) \\ &= \frac{1}{2}\log\left(2^{2h(Y_i|X_{2,i},Q_i)} + 2\pi e(N_2 - N)\right) \end{aligned}$$

where $(a)$ follows from Jensen's inequality and the fact that $\log(2^x + c)$ is a convex function.

Summing over the index $i$, the preceding inequality becomes
$$\begin{aligned} \sum_{i=1}^{n} h(Y_i + Z'_i|X_{2,i}, Q_i) &\geq \frac{1}{2}\sum_{i=1}^{n}\log\left(2^{2h(Y_i|X_{2,i},Q_i)} + 2\pi e(N_2 - N)\right) \\ &\stackrel{(a)}{\geq} \frac{n}{2}\log\left(2^{2\frac{1}{n}\sum_{i=1}^{n} h(Y_i|X_{2,i},Q_i)} + 2\pi e(N_2 - N)\right) \\ &\stackrel{(b)}{=} \frac{n}{2}\log\left(2\pi e(\alpha P_1 + N) + 2\pi e(N_2 - N)\right) \\ &= \frac{n}{2}\log\left(2\pi e(\alpha P_1 + N_2)\right) \end{aligned}$$

where (a) follows from Jensen's inequality, and (b) follows from (67).

Applying the preceding bound to the term $\sum_{i=1}^{n} h(Y_{2,i}|X_{2,i}, Q_i)$, we obtain

$$\sum_{i=1}^{n} h(Y_{2,i}|X_{2,i}, Q_i) = \sum_{i=1}^{n} h(Y_i + Z'_i|X_{2,i}, Q_i) \geq \frac{n}{2} \log\left(2\pi e(\alpha P_1 + N_2)\right) \quad (72)$$

Substituting the preceding bound into (71), we obtain

$$\begin{aligned} nR_{1,e} &\leq \frac{n}{2} \log\left(1 + \frac{\alpha P_1}{N}\right) - \frac{n}{2} \log\left(2\pi e(\alpha P_1 + N_2)\right) + \frac{n}{2} \log 2\pi e N_2 + n\delta_n \\ &\leq \frac{n}{2} \log\left(1 + \frac{\alpha P_1}{N}\right) - \frac{n}{2} \log\left(1 + \frac{\alpha P_1}{N_2}\right) + n\delta_n \end{aligned} \quad (73)$$

From (31), we obtain

$$nR_0 + nR_{1,e}$$
$$\leq \sum_{i=1}^{n} I(X_{1,i}, X_{2,i}; Y_i) - I(X_{1,i}; Y_{2,i}|X_{2,i}, Q_i) + n\delta_n$$
$$\leq \frac{n}{2} \log\left(1 + \frac{P_1 + P_2 + 2\sqrt{\bar{\alpha} P_1 P_2}}{N}\right) - \sum_{i=1}^{n} h(Y_{2,i}|X_{2,i}, Q_i) + \frac{n}{2} \log 2\pi e N_2 + n\delta_n \quad (74)$$
$$\leq \frac{n}{2} \log\left(1 + \frac{P_1 + P_2 + 2\sqrt{\bar{\alpha} P_1 P_2}}{N}\right) - \frac{n}{2} \log\left(2\pi e(\alpha P_1 + N_2)\right) + \frac{n}{2} \log 2\pi e N_2 + n\delta_n$$
$$\leq \frac{n}{2} \log\left(1 + \frac{P_1 + P_2 + 2\sqrt{\bar{\alpha} P_1 P_2}}{N}\right) - \frac{n}{2} \log\left(1 + \frac{\alpha P_1}{N_2}\right) + n\delta_n,$$

which completes the proof. □

## 6 GMAC with Two Confidential Message Sets

In this section, we consider the general case of the GMAC with (two) confidential messages, where the two users have common messages and each user has private (confidential) messages intended for the destination. Each user wants to keep the other user as ignorant of its confidential messages as possible. For this case, the rate-equivocation tuple has five components and takes the form $(R_0, R_1, R_2, R_{1,e}, R_{2,e})$, where $R_{1,e}$ and $R_{2,e}$ are equivocation rates indicating the secrecy levels of confidential messages sent by user 1 and confidential messages sent by user 2, respectively.

In the following, we first provide our main results on the achievable rate-equivocation region and secrecy rate region. It can be seen that the rate-equivocation region has a more complicated form compared to that of the case with a single confidential message set, and carries a new feature of the trade-off between the secrecy levels that can be achieved for the two confidential message sets. We then give an intuitive interpretation of the rate-equivocation region, following which we provide a rigorous proof of the rate-equivocation region. Finally, we give an equivalence proof of two regions, which simplifies the rate-equivocation region to an explicit form.

## 6.1 Main Results

The following theorem provides an inner bound on the capacity-equivocation region for the GMAC with two confidential message sets.

**Theorem 7.** *The following convexified region of nonnegative rate-equivocation tuples is achievable for the GMAC with two confidential message sets:*

$$\mathscr{R}^{II} = \textbf{Convex} \bigcup_{\substack{p(q)p(u|q)p(x_1|u) \\ p(v|q)p(x_2|v) \\ p(y,y_1,y_2|x_1,x_2)}} \left\{ \begin{array}{l} (R_0, R_1, R_2, R_{1,e}, R_{2,e}): \\ R_0 \geq 0, R_1 \geq 0, R_2 \geq 0, \\ R_1 \leq I(U;Y|V,Q), \\ R_2 \leq I(V;Y|U,Q), \\ R_1 + R_2 \leq I(U,V;Y|Q), \\ R_0 + R_1 + R_2 \leq I(U,V,Q;Y), \\ (R_{1,e}, R_{2,e}) \in \mathcal{S}_e(R_0, R_1, R_2) \end{array} \right\}. \quad (75)$$

*where*

$$\mathcal{S}_e(R_0, R_1, R_2) = \bigcup_{\substack{(R'_1, R'_2): \\ (R_0, R'_1, R'_2) \in \mathcal{C}^p_{MAC}, \\ R_1 \leq R'_1, R_2 \leq R'_2}} \left\{ \begin{array}{l} (R_0, R_{1,e}, R_{2,e}): \\ 0 \leq R_{1,e} \leq R_1 \\ R_{1,e} \leq [R'_1 - I(U;Y_2|X_2, V, Q)]_+ \\ 0 \leq R_{2,e} \leq R_2 \\ R_{2,e} \leq [R'_2 - I(V;Y_1|X_1, U, Q)]_+ \end{array} \right\} \quad (76)$$

*where $\mathcal{C}^p_{MAC}$ is defined as*

$$\mathcal{C}^p_{MAC} = \left\{ \begin{array}{l} (R_0, R_1, R_2): \\ R_0 \geq 0, R_1 \geq 0, R_2 \geq 0, \\ R_1 \leq I(U;Y|V,Q), \\ R_2 \leq I(V;Y|U,Q), \\ R_1 + R_2 \leq I(U,V;Y|Q), \\ R_0 + R_1 + R_2 \leq I(U,V,Q;Y) \end{array} \right\}. \quad (77)$$

An intuitive interpretation of the rate-equivocation region in Theorem 7 is given in the next subsection. The proof will be provided in Section 6.3.

Note that the region $\mathcal{S}_e(R_0, R_1, R_2)$ contains the secrecy rate pairs $(R_{1,e}, R_{2,e})$ that can be achieved for the given rate tuple $(R_0, R_1, R_2)$. It can be seen that $\mathcal{S}_e(R_0, R_1, R_2)$ in (76) is characterized by a union of a set of $(R'_1, R'_2)$, which is not an explicit form. It is thus desirable to change $\mathcal{S}_e(R_0, R_1, R_2)$ to an explicit form that is characterized through inequality bounds only.

**Theorem 8.** *The set $\mathcal{S}_e(R_0, R_1, R_2)$ in (76) can be expressed in the following explicit form:*

$$\mathcal{S}_e(R_0, R_1, R_2) = \mathfrak{L}_1(R_0, R_1, R_2) \cup \mathfrak{L}_2(R_0, R_1, R_2) \cup \mathfrak{L}_3(R_0, R_1, R_2) \tag{78}$$

*where*

$$\mathfrak{L}_1(R_0, R_1, R_2) = \left\{ \begin{array}{l} (R_{1,e}, R_{2,e}): \\ 0 \leq R_{1,e} \leq R_1 \\ 0 \leq R_{2,e} \leq R_2 \\ R_{1,e} \leq [I(U;Y|V,Q) - I(U;Y_2|X_2,V,Q)]_+, \\ R_{1,e} \leq [I(U,V;Y|Q) - R_2 - I(U;Y_2|X_2,V,Q)]_+, \\ R_{1,e} \leq [I(U,V,Q;Y) - R_0 - R_2 - I(U;Y_2|X_2,V,Q)]_+ \\ R_{2,e} \leq [I(V;Y|U,Q) - I(V;Y_1|X_1,U,Q)]_+, \\ R_{2,e} \leq [I(U,V;Y|Q) - R_1 - I(V;Y_1|X_1,U,Q)]_+, \\ R_{2,e} \leq [I(U,V,Q;Y) - R_0 - R_1 - I(V;Y_1|X_1,U,Q)]_+ \\ R_{1,e} + R_{2,e} \leq [I(U,V;Y|Q) - I(U;Y_2|X_2,V,Q) - I(V;Y_1|X_1,U,Q)]_+, \\ R_{1,e} + R_{2,e} \leq [I(U,V,Q;Y) - R_0 - I(U;Y_2|X_2,V,Q) - I(V;Y_1|X_1,U,Q)]_+ \end{array} \right\} \tag{79}$$

$$\mathfrak{L}_2(R_0, R_1, R_2) = \left\{ \begin{array}{l} (R_{1,e}, R_{2,e}): \\ 0 \leq R_{1,e} \leq R_1 \\ R_{1,e} \leq [I(U;Y|V,Q) - I(U;Y_2|X_2,V,Q)]_+, \\ R_{1,e} \leq [I(U,V;Y|Q) - R_2 - I(U;Y_2|X_2,V,Q)]_+, \\ R_{1,e} \leq [I(U,V,Q;Y) - R_0 - R_2 - I(U;Y_2|X_2,V,Q)]_+ \\ R_{2,e} = 0 \end{array} \right\} \tag{80}$$

*and*

$$\mathfrak{L}_3(R_0, R_1, R_2) = \left\{ \begin{array}{l} (R_{1,e}, R_{2,e}): \\ R_{1,e} = 0, \\ 0 \leq R_{2,e} \leq R_2 \\ R_{2,e} \leq [I(V;Y|U,Q) - I(V;Y_1|X_1,U,Q)]_+, \\ R_{2,e} \leq [I(U,V;Y|Q) - R_1 - I(V;Y_1|X_1,U,Q)]_+, \\ R_{2,e} \leq [I(U,V,Q;Y) - R_0 - R_1 - I(V;Y_1|X_1,U,Q)]_+ \end{array} \right\} \tag{81}$$

The proof of Theorem 8 is given in Section 6.4.

**Remark 10.** *The region (75) reduces to the capacity region of the MAC [15] by removing the secrecy constraints ($R_{1,e} = 0, R_{2,e} = 0$) and setting $U = X_1$ and $V = X_2$.*

**Remark 11.** *The last two bounds in (79) indicate that there is a trade-off between the two equivocation rates $R_{1,e}$ and $R_{2,e}$, i.e., the secrecy levels achieved for the two confidential* message sets $W_1$ and $W_2$. It will be clear in Section 6.2 that this trade-off can be achieved by using codebooks that achieve different boundary points of the MAC. It will be further clear in Section 6.3 that this trade-off corresponds to a trade-off between the sizes of the two codebooks used by the two users.

**Remark 12.** *The sets $\mathfrak{L}_2(R_0, R_1, R_2)$ and $\mathfrak{L}_3(R_0, R_1, R_2)$ characterize the equivocation rates when only user 1 or user 2 achieves nonzero equivocation rates.*

We now study the case where confidential messages of each user are perfectly hidden from the other user. This happens when $R_{1,e} = R_1$ and $R_{2,e} = R_2$. The rate region that contains all these rate tuples is called the *secrecy capacity region* and is given by

$$\mathcal{C}_s^{II} = \{(R_0, R_1, R_2) : (R_0, R_1, R_2, R_1, R_2) \in \mathscr{C}^{II}\}. \tag{82}$$

The following inner bound on the secrecy capacity region $\mathcal{C}_s^{II}$ follows from Theorem 7 and Theorem 8.

**Corollary 4.** *A secrecy rate region (inner bound on secrecy capacity region) for the GMAC with two confidential message sets is given by:*

$$\mathcal{R}_s^{II} = \mathbf{Convex} \bigcup_{\substack{p(q)p(u|q)p(x_1|u) \\ p(v|q)p(x_2|v)p(y,y_1,y_2|x_1,x_2)}} \left\{ \mathcal{R}_{s,1} \cup \mathcal{R}_{s,2} \cup \mathcal{R}_{s,3} \right\} \tag{83}$$

*where*

$$\mathcal{R}_{s,1} = \left\{ \begin{array}{l} (R_0, R_1, R_2): \\ R_0 \geq 0, R_1 \geq 0, R_2 \geq 0, \\ R_1 \leq I(U;Y|V,Q) - I(U;Y_2|V,X_2,Q), \\ R_2 \leq I(V;Y|U,Q) - I(V;Y_1|U,X_1,Q), \\ R_1 + R_2 \leq I(U,V;Y|Q) - I(U;Y_2|V,X_2,Q) - I(V;Y_1|U,X_1,Q), \\ R_0 + R_1 + R_2 \leq I(U,V,Q;Y) - I(U;Y_2|V,X_2,Q) - I(V;Y_1|U,X_1,Q) \end{array} \right\} \tag{84}$$

$$\mathcal{R}_{s,2} = \left\{ \begin{array}{l} (R_0, R_1, R_2): \\ R_0 \geq 0, R_1 \geq 0, R_2 = 0, \\ R_1 \leq I(U;Y|V,Q) - I(U;Y_2|V,X_2,Q), \\ R_0 + R_1 \leq I(U,V,Q;Y) - I(U;Y_2|V,X_2,Q) \end{array} \right\} \tag{85}$$

$$\mathcal{R}_{s,3} = \left\{ \begin{array}{l} (R_0, R_1, R_2): \\ R_0 \geq 0, R_2 \geq 0, R_1 = 0, \\ R_2 \leq I(V;Y|U,Q) - I(V;Y_1|U,X_1,Q), \\ R_0 + R_2 \leq I(U,V,Q;Y) - I(V;Y_1|U,X_1,Q) \end{array} \right\} \tag{86}$$

For the GMAC with two confidential message sets, it is desirable that both users achieve positive secrecy rates. We have the following sufficient condition for this case to happen.

**Corollary 5.** *A sufficient condition for both users to have positive secrecy rates is that*

$$I(U;Y|V,Q) > I(U;Y_2|X_2,V,Q) \qquad (87)$$

*for some joint distribution $p(q)p(u|q)p(x_1|u)p(v|q)p(x_2|v)p(y,y_1,y_2|x_1,x_2)$, and*

$$I(V;Y|U,Q) > I(V;Y_1|X_1,U,Q) \qquad (88)$$

*for some joint distribution $p(q)p(u|q)p(x_1|u)p(v|q)p(x_2|v)p(y,y_1,y_2|x_1,x_2)$.*

Note that the joint distributions that satisfy the two conditions (87) and (88) are not necessarily the same.

*Proof.* From (85), it is clear that if (87) is satisfied, user 1 achieves positive secrecy rate (for $R_0 = 0$). Similarly, from (86), user 2 achieves positive rate if (88) is satisfied. Therefore, time-sharing between these two operating points guarantees positive secrecy rates for both users. □

In Fig. 8, we plot the secrecy rate region for the GMAC with two confidential message sets for a given distribution $p(q)p(u|q)p(x_1|u)p(v|q)p(x_2|v)p(y,y_1,y_2|x_1,x_2)$. For the given joint distribution, we assume that both conditions in Corollary 5 are satisfied, i.e., both users have nonzero secrecy rates. Moreover, we assume that the bounds on the sum rate $R_1 + R_2$ in (84) is positive, i.e., the region $\mathcal{R}_{s,1}$ is nonempty. The geometric structure of the secrecy rate region (shaded area) falls into one of the four cases depending on how the mutual information terms compare with each other. In Fig. 8 we also plot the capacity of the corresponding MAC without secrecy constraints (setting $Y_1 = \phi$ and $Y_2 = \phi$ in the GMAC) with the outer solid line as the boundary. It is clear from the figure that the secrecy rate region is inside the capacity region of the corresponding MAC. Hence to achieve perfectly secure communication for confidential messages, the users need to transmit their confidential messages at smaller rates than reliable communication rates. It is also clear from Fig. 8 that the secrecy rate region given by $\mathcal{R}_{s,1} \cup \mathcal{R}_{s,2} \cup \mathcal{R}_{s,3}$ is not convex in general and needs to be convexfied. These cases of secrecy regions will be further discussed in the next section and intuition behind the achievability of the corner points will be given.

## 6.2 Interpretation of $\mathscr{R}^{II}$ in Theorem 7

In this subsection, we explain the intuition behind the achievable rate-equivocation region $\mathscr{R}^{II}$ given in Theorem 7. A rigorous proof is relegated to the next subsection.

We focus on the following region $\tilde{\tilde{\mathscr{R}}}$ given in Theorem 7. The region $\mathscr{R}^{II}$ given in (75) follows from $\tilde{\tilde{\mathscr{R}}}$ by prefixing two discrete memoryless channels with inputs $U$ and $V$ and

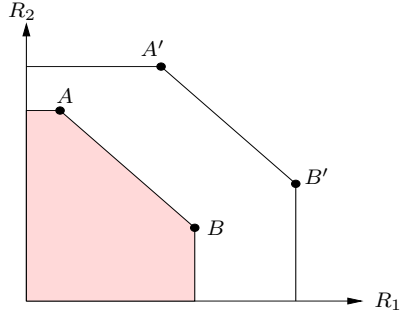

Case 1: $I(V;Y|Q) > I(V;Y_1|X_1,U,Q)$ and $I(U;Y|Q) > I(U;Y_2|X_2,V,Q)$

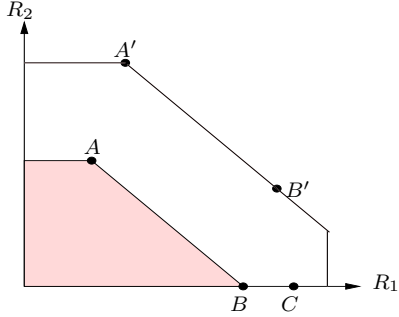
(a) before convexification

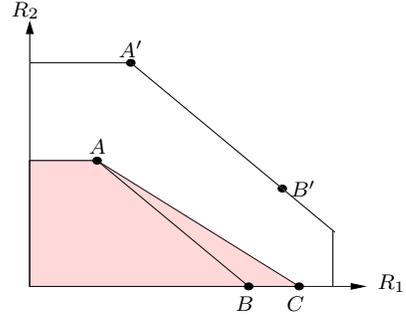
(b) after convexification

Case 2: $I(V;Y|Q) \leq I(V;Y_1|X_1,U,Q)$ and $I(U;Y|Q) > I(U;Y_2|X_2,V,Q)$

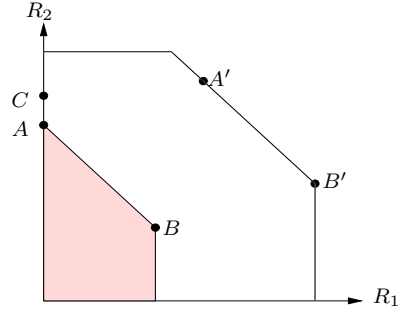
(a): before convexification

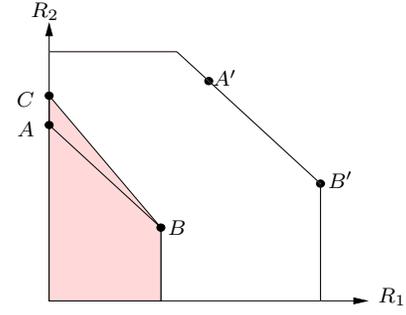
(b): after convexification

Case 3: $I(V;Y|Q) > I(V;Y_1|X_1,U,Q)$ and $I(U;Y|Q) \leq I(U;Y_2|X_2,V,Q)$

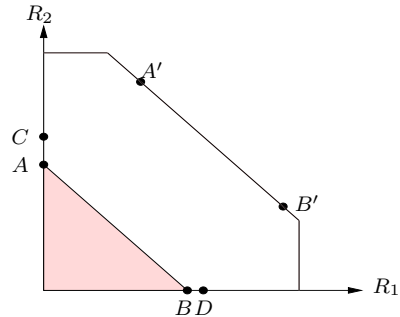
(a): before convexification

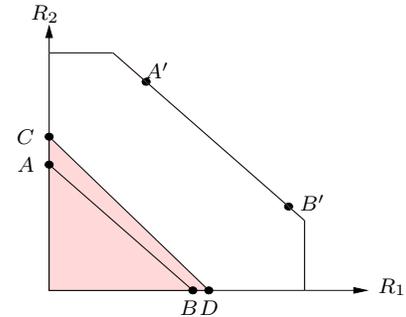
(b): after convexification

Case 4: $I(V;Y|Q) \leq I(V;Y_1|X_1,U,Q)$ and $I(U;Y|Q) \leq I(U;Y_2|X_2,V,Q)$

Figure 8: Secrecy rate region for the GMAC with two confidential message sets for a given joint distribution

transition probabilities $p(x_1|u)$ and $p(x_2|v)$ to the given GMAC (similar to [3, Lemma 4]).

$$\tilde{\mathscr{R}} = \bigcup_{\substack{p(q)p(x_1|q)p(x_2|q) \\ p(y,y_1,y_2|x_1,x_2)}} \left\{ \begin{array}{l} (R_0, R_1, R_2, R_{1,e}, R_{2,e}) : \\ R_1 \leq I(X_1; Y|X_2, Q), \\ R_2 \leq I(X_2; Y|X_1, Q), \\ R_1 + R_2 \leq I(X_1, X_2; Y|Q), \\ R_0 + R_1 + R_2 \leq I(X_1, X_2; Y), \\ (R_{1,e}, R_{2,e}) \in \tilde{\mathcal{S}}_e(R_0, R_1, R_2) \end{array} \right\}. \tag{89}$$

where

$$\tilde{\mathcal{S}}_e(R_0, R_1, R_2) = \bigcup_{(R'_1, R'_2) \in \mathcal{A}(R_0, R_1, R_2)} \left\{ \begin{array}{l} (R_{1,e}, R_{2,e}) : \\ 0 \leq R_{1,e} \leq R_1 \\ R_{1,e} \leq [R'_1 - I(X_1; Y_2|X_2, Q)]_+ \\ 0 \leq R_{2,e} \leq R_2 \\ R_{2,e} \leq [R'_2 - I(X_2; Y_1|X_1, Q)]_+ \end{array} \right\} \tag{90}$$

where

$$\mathcal{A}(R_0, R_1, R_2) := \{(R'_1, R'_2) : (R_0, R'_1, R'_2) \in \tilde{\mathcal{C}}^p_{MAC}, R_1 \leq R'_1, R_2 \leq R'_2\}, \tag{91}$$

and

$$\tilde{\mathcal{C}}^p_{MAC} = \left\{ \begin{array}{l} (R_0, R_1, R_2) : \\ R_1 \leq I(X_1; Y|X_2, Q), \\ R_2 \leq I(X_2; Y|X_1, Q), \\ R_1 + R_2 \leq I(X_1, X_2; Y|Q), \\ R_0 + R_1 + R_2 \leq I(X_1, X_2; Y) \end{array} \right\} \tag{92}$$

It is easy to see that the first four bounds in $\tilde{\mathscr{R}}$ in (89) are the bounds that define the capacity region of the MAC [15]. Hence these bounds also need to be satisfied for the GMAC with two confidential message sets. To understand the bounds on $(R_{1,e}, R_{2,e})$, we plot Fig. 9 for illustration. In Fig. 9, the solid line indicates the boundary of the region $\tilde{\mathcal{C}}^p_{MAC}$ for a given common rate $R_0$.

For a given rate tuple $(R_0, R_1, R_2)$ that satisfies the first four bounds in (89), which also means that $(R_0, R_1, R_2) \in \tilde{\mathcal{C}}^p_{MAC}$, we plot the region $\mathcal{A}(R_0, R_1, R_2)$ of $(R'_1, R'_2)$ in Fig. 9 as the shaded area. It is clear that the rate tuple $(R_0, R_1, R_2)$ can be achieved by applying the codebook that achieves any rate tuples $(R_0, R'_1, R'_2) \in \mathcal{A}(R_0, R_1, R_2)$ and throwing away the redundant bits $R'_1 - R_1$ and $R'_2 - R_2$. However, it is a waste of channel resources to achieve a rate tuple with lower rate components by applying a codebook that achieves a rate tuple with higher rate components.

The situation becomes different for channels with confidential message sets, where one user (say user 1) wants to keep its confidential information secret from the other user (user 2). Now to achieve a rate $R_1$, user 1 may transmit at a higher rate $R'_1$ so that the codebook contains a larger number of codewords than the number of messages that user 1 wants to convey to the destination. The redundant codewords are used to confuse user 2 about real messages that user 1 transmits to the destination. Since user 2 can decode at the rate of the capacity of the channel from user 1 to user 2, which is $I(X_1; Y_2|X_2, Q)$, intuitively user 1 can keep information with the rate $R'_1 - I(X_1; Y_2|X_2, Q)$ secret from user 2. Similarly, user 2 can keep information with the rate $R'_2 - I(X_2; Y_1|X_1, Q)$ secret from user 1. Hence the equivocation rates $R_{1,e} = R'_1 - I(X_1; Y_2|X_2, Q)$ and $R_{2,e} = R'_2 - I(X_2; Y_1|X_1, Q)$ can be achieved by the codebook achieving $(R_0, R'_1, R'_2)$ for the MAC. We hence conclude that all equivocation rate pair $(R_{1,e}, R_{2,e})$ in $\tilde{\mathcal{S}}_e(R_0, R_1, R_2)$ given in (90) are achievable.

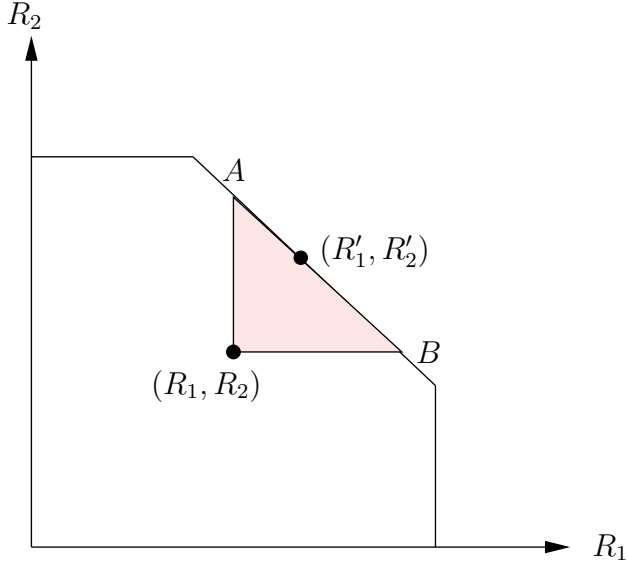

Figure 9: Region $\tilde{\mathcal{C}}^p_{MAC}$ for a given common rate $R_0$

Note that there is no loss of generality to consider only the rate tuples $(R_0, R'_1, R'_2) \in \mathcal{A}(R_0, R_1, R_2)$ that are on the sum rate boundary of the region $\tilde{\mathcal{C}}^p_{MAC}$, i.e., the points on the line between the point $A$ and point $B$ in Fig. 9. This is because any point inside $\mathcal{A}(R_0, R_1, R_2)$ corresponds to a point on the line from $A$ to $B$ that achieves larger $R_{1,e}$ and $R_{2,e}$. To operate at different points between $A$ and $B$, user 1 and user 2 use different sizes of codebooks, and achieve different equivocation rate pairs $(R_{1,e}, R_{2,e})$. As the operating point moves from $A$ to $B$, the equivocation rate $R_{1,e}$ increases and $R_{2,e}$ decreases, thus achieving a trade-off between the security levels for the two confidential message sets sent by user 1 and user 2 as commented in Remark 11.

Based on the preceding interpretation, we next give intuition for the secrecy rate region given in Corollary 4. In particular, we focus on the secrecy rate region of case 1 in Fig. 8. We plot this case in more detail in Fig. 10.

We first consider case 1, where both of the following two conditions are satisfied:

$$I(V; Y|Q) > I(V; Y_1|X_1, U, Q) \qquad (93)$$

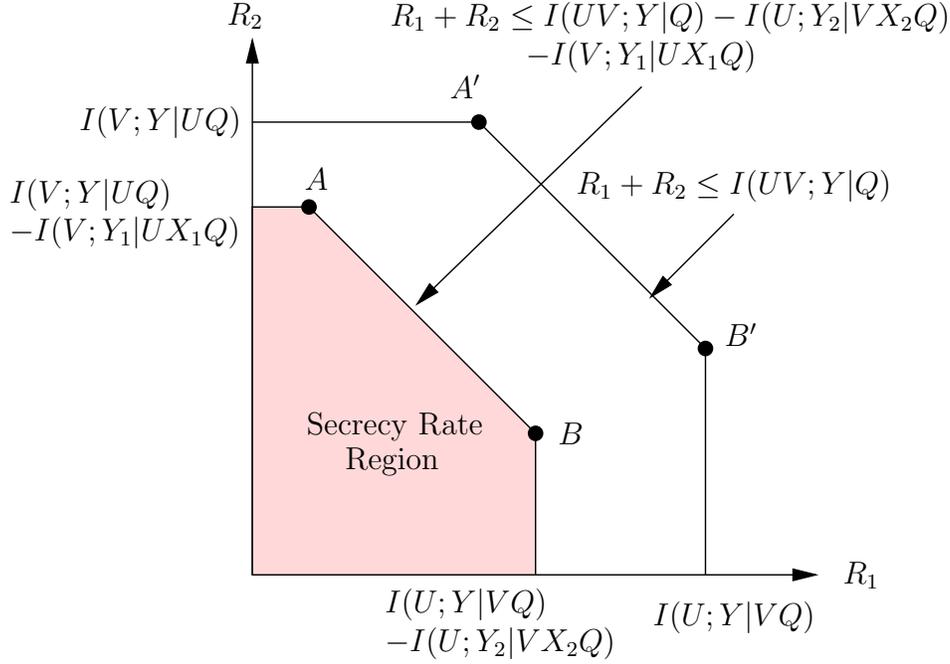

Figure 10: Secrecy rate region of case 1: $I(V;Y|Q) > I(V;Y_1|X_1,U,Q)$ and $I(U;Y|Q) > I(U;Y_2|X_2,V,Q)$

and
$$I(U;Y|Q) > I(U;Y_2|X_2,V,Q) \tag{94}$$

so that the secrecy rate region is a pentagon. All rate tuples in the secrecy rate region can be achieved with confidential messages of the two users being perfectly secret from each other. However, to achieve perfect secrecy for rate tuples in the shaded area, the two users may need to use codebooks achieving rate tuples outside the shaded area with larger rate components. In this way, the size of the codebook used by each user is larger than the actual number of confidential messages that need to be delivered, and the redundant codewords in the codebook are used to confuse the other user. For example, consider the corner point $A$ of the secrecy rate region in Fig. 10. The rates of the two confidential message sets are:

$$R_{A,1} = I(U;Y|Q) - I(U;Y_2|V,X_2,Q) \tag{95}$$

$$R_{A,2} = I(V;Y|U,Q) - I(V;Y_1|U,X_1,Q) \tag{96}$$

To achieve the point $A$ with perfect secrecy, the actual codebook being used needs to operate at the point $A'$ which is on the boundary of the capacity region of the MAC. The rates corresponding to $A'$ are:
$$R_{A',1} = I(U;Y|Q) \tag{97}$$

$$R_{A',2} = I(V;Y|U,Q) \tag{98}$$

If we compare the rate tuples corresponding to the points $A$ and $A'$, it is easy to see that the secrecy rates decrease from the actual rates of the codebook by the capacities of the channels between the two users, i.e., $I(U; Y_2|V, X_2, Q)$ and $I(V; Y_1|U, X_1, Q)$. In fact, every point on the boundary of the secrecy rate region of case 1 in Fig. 10 is achieved by the codebook that operates at a corresponding point on the boundary of the capacity region of the MAC.

## 6.3 Proof of the Achievability of $\mathscr{R}^{II}$ in Theorem 7

In this section, we show that the rate-equivocation region $\mathscr{R}^{II}$ given in Theorem 7 is achievable. We first show that the region $\tilde{\mathscr{R}}$ given in (89) is achievable. Then $\mathscr{R}^{II}$ in Theorem 7 is achievable follows by prefixing two discrete memoryless channels as reasoned at the beginning of the preceding section.

We present the proof in four steps. In Step 1, we prove existence of a certain codebook based on a random coding technique, which is different from the nonrandom code construction used in [3]. Our random coding proof is also different from the proof in [2] in that the codeword ensemble contains only typical sequences, which makes equivocation computation convenient. In Step 2, we define our encoding scheme. In Step 3, we compute the two equivocation rates. The technique follows [3, Sec. IV]. In Step 4, we consider other cases where the encoding scheme is slightly different from the case considered in the preceding steps.

**Step 1: Existence of Certain Codebook**

We consider the following joint distribution

$$P_{QX_1X_2YY_1Y_2} = p(q)p(x_1|q)p(x_2|q)p(y, y_1, y_2|x_1, x_2)$$

We use $T_\epsilon^n(P_{QX_1X_2YY_1Y_2})$ to indicate the strong typical set defined by the distribution $P_{QX_1X_2YY_1Y_2}$.

Consider a given rate triple $(R_0, R_1, R_2) \in \tilde{\mathcal{C}}_{MAC}^p$, where $\tilde{\mathcal{C}}_{MAC}^p$ is given in (92). We wish to find a codebook that achieves $(R_0, R_1, R_2)$ with small probability of error, and achieves certain equivocation rates $R_{1,e}$ and $R_{2,e}$. We consider $(R_1', R_2')$ that satisfy the following conditions:

$$(R_0, R_1', R_2') \in \tilde{\mathcal{C}}_{MAC}^p, \qquad R_1 \leq R_1', \qquad R_2 \leq R_2' \tag{99}$$

and

$$R_1' - I(X_1; Y_2|X_2, Q) > 0, \qquad R_2' - I(X_2; Y_1|X_1, Q) > 0 \tag{100}$$

The cases where $R_1' - I(X_1; Y_2|X_2, Q) \leq 0$ or $R_2' - I(X_2; Y_1|X_1, Q) \leq 0$ will be considered in Step 4.

The following lemma states existence of a certain codebook, which will be used for encoding in Step 2.

**Lemma 5.** *For a given rate triple* $(R_0, R_1, R_2) \in \tilde{\mathcal{C}}^p_{MAC}$, *there exists the following codebook:*

$$\mathbb{C} = \begin{cases} q_i^n, & i = 1, \ldots, 2^{nR_0}; \\ x_{1,iab}^n, & i = 1, \ldots, 2^{nR_0}; a = 1, \ldots, A; b = 1, \ldots, B; \\ x_{2,ist}^n, & i = 1, \ldots, 2^{nR_0}; s = 1, \ldots, S; t = 1, \ldots, T; \end{cases} \quad (101)$$

*where all codewords are strongly typical, i.e.,* $q_i^n \in T_\epsilon^n(P_Q), x_{1,iab}^n \in T_\epsilon^n(P_{X_1|Q}|q_i^n), x_{2,ist}^n \in T_\epsilon^n(P_{X_2|Q}|q_i^n)$ *for all* $i, a, b, s, t$. *The number of codewords are defined as follows:*

$$\begin{aligned}
\frac{1}{n} \log A &= R_1' - I(X_1; Y_2 | X_2, Q), \\
\frac{1}{n} \log B &= I(X_1; Y_2 | X_2, Q), \\
\frac{1}{n} \log S &= R_2' - I(X_2; Y_1 | X_1, Q), \\
\frac{1}{n} \log T &= I(X_2; Y_1 | X_1, Q).
\end{aligned} \quad (102)$$

We define the following probabilities of error when the codewords $x_{1,iab}^n$ and $x_{2,ist}^n$ are transmitted by user 1 and user 2, respectively:

$$\begin{aligned}
\lambda_{iabst} &= \text{Error probability for the destination to decode } x_{1,iab}^n \text{ and } x_{2,ist}^n; \\
\lambda_{1,b|iast} &= \text{Error probability for user 2 to decode } x_{1,iab}^n \text{ given } i, a, s, t; \\
\lambda_{2,t|iabs} &= \text{Error probability for user 1 to decode } x_{2,ist}^n \text{ given } i, a, b, s.
\end{aligned} \quad (103)$$

Let $p_{iabst}$ be the probability that codewords $x_{1,iab}^n$ and $x_{2,ist}^n$ are transmitted by user 1 and user 2, respectively. We further define the following average probabilities of error:

$$\begin{aligned}
\lambda &= \sum_{iabst} p_{iabst} \lambda_{iabst}; \\
\lambda_1 &= \sum_{iabst} p_{iabst} \lambda_{1,b|iast}; \\
\lambda_2 &= \sum_{iabst} p_{iabst} \lambda_{2,t|iabs}.
\end{aligned} \quad (104)$$

For sufficiently large $n$, the codebook described in (101) satisfies

$$\lambda < \epsilon, \quad \lambda_1 < \epsilon, \quad \text{and} \quad \lambda_2 < \epsilon, \quad (105)$$

for any arbitrary $0 < \epsilon < 1$.

In Fig. 11, we plot an example codebook that is described in the preceding lemma. We can interpret each row in the codebook as a subcodebook. If each of the two users randomly chooses one codeword from its codebook and sends it over the channel, the destination can decode these pair of codewords with small average error probability, because $\lambda < \epsilon$.

### Codebook for user 1

| b \ a | 1 | 2 | 3 | ⋯ | ⋯ | $B = 2^{nI(X_1;Y_2|X_2Q)}$ |
|---|---|---|---|---|---|---|
| 1 | $x_{1,i11}^n$ | $x_{1,i12}^n$ | . | . | . | . |
| 2 | $x_{1,i21}^n$ | . | . | . | . | . |
| 3 | . | . | . | . | . | . |
| ⋮ | . | . | . | . | . | . |
| ⋮ | . | . | . | . | . | . |
| $A = 2^{n(R_1' - I(X_1;Y_2|X_2Q))}$ | . | . | . | . | . | . |

Codebook for user 1

### Codebook for user 2

| t \ s | 1 | 2 | 3 | ⋯ | ⋯ | $T = 2^{nI(X_2;Y_1|X_1Q)}$ |
|---|---|---|---|---|---|---|
| 1 | $x_{2,i11}^n$ | $x_{2,i12}^n$ | . | . | . | . |
| 2 | $x_{2,i21}^n$ | . | . | . | . | . |
| 3 | . | . | . | . | . | . |
| ⋮ | . | . | . | . | . | . |
| ⋮ | . | . | . | . | . | . |
| $S = 2^{n(R_2' - I(X_2;Y_1|X_1Q))}$ | . | . | . | . | . | . |

Codebook for user 2

Figure 11: Codebook for users 1 and 2

However, each user as a receiver can only decode the codeword sent by the other user with small average probability of error if it knows to which row the transmitted codeword belongs. This is because $\lambda_1 < \epsilon$ and $\lambda_2 < \epsilon$. Therefore, while the destination decodes reliably over the entire codebook, the two users decode reliably only within rows of the codebook.

*Proof.* We prove the lemma using a random coding technique. We define the following sum of error probabilities:

$$p_e = \lambda + \lambda_1 + \lambda_2 = \sum_{iabst} p_{iabst}(\lambda_{iabst} + \lambda_{1,b|iast} + \lambda_{2,t|iabs}). \tag{106}$$

We show that the average of $p_e$ over a random codebook ensemble is small for sufficiently large codeword length $n$. Then, there must exist at least one codebook such that $p_e$ is small for sufficiently large $n$.

For a given distribution $p(q)p(x_1|q)p(x_2|q)$, we construct random codebooks by the following generating steps.

1. Generate $2^{nR_0}$ codewords $q^n$, each uniformly drawn from the set $T_\epsilon^n(P_Q)$. Index $q_i^n$, $i = 1, \ldots, 2^{nR_0}$.

2. For each $q_i^n$, generate $2^{nR_1'}$ codewords $x_1^n$, each uniformly drawn from the set $T_\epsilon^n(P_{X_1|Q}|q_i^n)$. Index $x_{1,iab}^n$, $a = 1, \ldots, A; b = 1, \ldots, B$.

3. For each $q_i^n$, generate $2^{nR_2'}$ codewords $x_2^n$, each uniformly drawn from the set $T_\epsilon^n(P_{X_2|Q}|q_i^n)$. Index $x_{2,ist}^n$, $s = 1, \ldots, S; t = 1, \ldots, T$.

Suppose the codewords $x_{1,iab}^n$ and $x_{2,ist}^n$ are transmitted by user 1 and user 2, respectively, we define the following decoding strategies at the destination, user 1 and user 2.

1. Destination declares that the indices of $x_1^n$ and $x_2^n$ are $\hat{i}, \hat{a}, \hat{b}, \hat{s}, \hat{t}$ if there is a unique group of such indices such that $\left(q_{\hat{i}}^n, x_{1,\hat{i}\hat{a}\hat{b}}^n, x_{2,\hat{i}\hat{s}\hat{t}}^n, y^n\right) \in T_\epsilon^n(P_{QX_1X_2Y})$.

2. User 2, given $i, a, s, t$, declares that the index $b$ of $x_{1,iab}^n$ is $\hat{b}$ if there is a unique $\hat{b}$ such that $\left(q_i^n, x_{1,ia\hat{b}}^n, x_{2,ist}^n, y_2^n\right) \in T_\epsilon^n(P_{QX_1X_2Y_2})$.

3. User 1, given $i, a, b, s$, declares that the index $t$ of $x_{2,ist}^n$ is $\hat{t}$ if there is a unique $\hat{t}$ such that $\left(q_i^n, x_{1,iab}^n, x_{2,is\hat{t}}^n, y_1^n\right) \in T_\epsilon^n(P_{QX_1X_2Y_1})$.

We can compute $\mathrm{E}_\mathbb{C}[p_e]$ by following the standard techniques as in [15, Chap. 14], where $\mathrm{E}_\mathbb{C}$ indicates averaging over the random codebook ensemble. We can show that

$$\mathrm{E}_\mathbb{C}[p_e] < \epsilon, \tag{107}$$

for sufficiently large codeword length $n$, by using the fact that $(R_0, R_1', R_2') \in \tilde{\mathcal{C}}_{MAC}^p$ and the sizes of indices $b$ and $t$ are $B = 2^{nI(X_1;Y_2|X_2,Q)}$ and $T = 2^{nI(X_2;Y_1|X_1,Q)}$, respectively.

Hence there exists one codebook such that for sufficiently large codebook size $n$,

$$p_e = \lambda + \lambda_1 + \lambda_2 < \epsilon. \tag{108}$$

This leads to the conclusion that for sufficiently large codebook size $n$

$$\lambda < \epsilon, \qquad \lambda_1 < \epsilon, \qquad \text{and} \qquad \lambda_2 < \epsilon. \tag{109}$$

$\square$

### Step 2: Encoding

Based on the codebook given in Lemma 5, we define an encoding strategy to achieve the given rate tuple $(R_0, R_1, R_2)$ with certain equivocation rates $(R_{1,e}, R_{2,e})$. The equivocation rates $(R_{1,e}, R_{2,e})$ will be computed in Step 3.

We first assume that $R_1 > R_1' - I(X_1; Y_2|X_2, Q)$ and $R_2 > R_2' - I(X_2; Y_1|X_1, Q)$. The cases where $R_1 \leq R_1' - I(X_1; Y_2|X_2, Q)$ or $R_2 \leq R_2' - I(X_2; Y_1|X_1, Q)$ will be considered in Step 4.

We denote common messages by $W_0 \in \mathcal{W} = [1, 2^{nR_0}]$, and denote confidential messages sent by the two users by $W_1 \in \mathcal{W}_1 = [1, 2^{nR_1}]$ and $W_2 \in \mathcal{W}_2 = [1, 2^{nR_2}]$, respectively. We further define the following sets

$$\mathcal{A} = [1, A], \ \mathcal{B} = [1, B], \ \mathcal{S} = [1, S], \ \mathcal{T} = [1, T] \tag{110}$$

where $A, B, S, T$ are defined in (102). We let

$$\mathcal{W}_1 = \mathcal{A} \times \mathcal{J} \tag{111}$$

where $\mathcal{J} = [1, J]$ and $\frac{1}{n} \log J = R_1 - [R_1' - I(X_1; Y_2|X_2, Q)]$, and

$$\mathcal{W}_2 = \mathcal{S} \times \mathcal{K} \tag{112}$$

where $\mathcal{K} = [1, K]$ and $\frac{1}{n} \log K = R_2 - [R_2' - I(X_2; Y_1|X_1, Q)]$.

We define the following mappings:

$$\begin{aligned} g_1 &: \mathcal{B} \to \mathcal{J}, \text{ partitioning } \mathcal{B} \text{ into } J \text{ subsets with } \textit{nearly equal size}; \\ g_2 &: \mathcal{T} \to \mathcal{K}, \text{ partitioning } \mathcal{T} \text{ into } K \text{ subsets with } \textit{nearly equal size}, \end{aligned} \tag{113}$$

where "early equal size" means

$$\|g_1^{-1}(j_1)\| \leq 2\|g_1^{-1}(j_2)\| \qquad \forall\, j_1, j_2 \in \mathcal{J} \tag{114}$$

and

$$\|g_2^{-1}(k_1)\| \leq 2\|g_2^{-1}(k_2)\| \qquad \forall\, k_1, k_2 \in \mathcal{K}. \tag{115}$$

The two encoders at users 1 and 2 are defined in the following:

$$\begin{aligned} f_0 &: \ W_0 \to \{q_i^n\}, \text{ mapping } w_0 \to i; \\ f_1 &: \ (W_0, W_1) \to \{x_{1,iab}^n\}, \text{ mapping } w_0 \to i, \text{ and mapping } w_1 = (a, j) \to (a, b), \\ &\quad \text{where } b \text{ is chosen uniformly from the set } g_1^{-1}(j) \subset \mathcal{B}; \\ f_2 &: \ (W_0, W_2) \to \{x_{2,ist}^n\}, \text{ mapping } w_0 \to i, \text{ and mapping } w_2 = (s, k) \to (s, t), \\ &\quad \text{where } t \text{ is chosen uniformly from the set } g_2^{-1}(k) \subset \mathcal{T}. \end{aligned} \tag{116}$$

The idea of the above encoding strategy is as follows. From Step 1, it is clear that users decode reliably within rows of the codebook and are not able to decode across different rows. Hence each user tries to map its confidential messages across different rows of the codebook to prevent the other user from decoding its messages.

**Step 3: Equivocation Computation**

Based on the codebook given in Lemma 5 in Step 1 and the encoding functions defined in (116) in Step 2, we have the following joint probability distribution:

$$p(w_0)p(w_1)p(w_2)f_0(q^n|w_0)f_1(x_1^n|w_0, w_1)f_2(x_2^n|w_0, w_2)p(y^n, y_1^n, y_2^n|x_1^n, x_2^n). \tag{117}$$

In (117), $p(w_0), p(w_1)$ and $p(w_2)$ are uniform distributions, i.e., the messages are uniformly chosen from the three message sets. The encoding function $f_0$ is a deterministic one-to-one mapping, and $f_1$ and $f_2$ are random mapping functions as defined in (116) in Step 2. For the joint distribution given in (117), we note the following two Markov chain conditions:

$$W_2 \to (X_2^n, W_0) \to (W_1, X_1^n, Y_1^n, Y_2^n, Y^n); \tag{118}$$

and

$$W_1 \to (X_1^n, W_0) \to (W_2, X_2^n, Y_1^n, Y_2^n, Y^n). \tag{119}$$

We first compute the equivocation rate of $W_1$ at user 2 in the following.

$$\begin{aligned}
H(W_1|Y_2^n, X_2^n, W_0, W_2) &\stackrel{(a)}{=} H(W_1|Y_2^n, X_2^n, W_0) \\
&= H(W_1, Y_2^n|X_2^n, W_0) - H(Y_2^n|X_2^n, W_0) \\
&= H(W_1, Y_2^n, X_1^n|X_2^n, W_0) - H(X_1^n|W_0, W_1, Y_2^n, X_2^n) - H(Y_2^n|X_2^n, W_0) \\
&= H(W_1, X_1^n|X_2^n, W_0) + H(Y_2^n|X_1^n, X_2^n, W_0, W_1) - H(X_1^n|W_0, W_1, Y_2^n, X_2^n) \\
&\quad - H(Y_2^n|X_2^n, W_0) \\
&\stackrel{(b)}{\geq} H(X_1^n|W_0) + H(Y_2^n|X_1^n, X_2^n) - H(X_1^n|W_0, W_1, Y_2^n, X_2^n) - H(Y_2^n|X_2^n, W_0)
\end{aligned} \tag{120}$$

In the preceding equation, $(a)$ follows from the Markov condition given in (118). The first term in $(b)$ follows from the following

$$\begin{aligned}
H(W_1, X_1^n|X_2^n, W_0) &= H(X_1^n|X_2^n, W_0) + H(W_1|X_1^n, X_2^n, W_0) \\
&\geq H(X_1^n|X_2^n, W_0) = H(X_1^n|W_0)
\end{aligned} \tag{121}$$

The second term in $(b)$ follows from the fact that $Y_2^n$ is independent of $W_0, W_1$ given $X_1^n, X_2^n$.

We now compute the four terms in (120) one by one. To compute the first term, we first show the following useful lemma.

**Lemma 6.** ([3]) *Consider a discrete random variable $X$ with the mass points being $x_1, \ldots, x_m$ and the probability mass function satisfying*

$$\frac{Pr\{X = x_i\}}{Pr\{X = x_j\}} \leq 2 \cdot 2^\delta \qquad \forall\, i, j \in [1, \ldots, m]. \tag{122}$$

*Then,*

$$H(X) \geq \log m - \delta - 1 \tag{123}$$

*Proof.* Let $p_i = Pr\{X = x_i\}$ for $i = 1, \ldots, m$. Let

$$p_{max} = \max\{p_1, \ldots, p_m\}, \quad \text{and} \quad p_{min} = \min\{p_1, \ldots, p_m\}. \tag{124}$$

We have $\frac{p_{max}}{p_{min}} \leq 2 \cdot 2^\delta$ by assumption. Hence

$$p_{max} \leq \frac{2 \cdot 2^\delta}{m} m p_{min} \leq \frac{2 \cdot 2^\delta}{m} \tag{125}$$

where we have used $mp_{min} \leq 1$. We can then bound the entropy of $X$ as follows.

$$H(X) = \sum_{i=1}^{m} -p_i \log p_i \geq \sum_{i=1}^{m} -p_i \log p_{max} \tag{126}$$
$$= -\log p_{max} \geq -\log \frac{2 \cdot 2^\delta}{m} = \log m - \delta - 1$$

□

For the first term in (120), we note that for each $W_0 = i$, $X_1^n$ has $2^{nR_1'}$ possible values. According to the encoding mapping function $f_1$ defined in (116), we have

$$\frac{Pr\{X_1^n = x_1^n\}}{Pr\{X_1^n = \bar{x}_1^n\}} \leq 2 \qquad \forall\, x_1^n, \bar{x}_1^n \in \{x_{1,iab}^n\} \tag{127}$$

By using Lemma 6, we obtain

$$\frac{1}{n} H(X_1^n | W_0) \geq R_1' - \frac{1}{n}. \tag{128}$$

For the second term in (120), we have

$$\frac{1}{n} H(Y_2^n | X_1^n, X_2^n)$$
$$= \frac{1}{n} \sum_{x_1^n, x_2^n} Pr\{X_1^n = x_1^n, X_2^n = x_2^n\} H(Y_2^n | X_1^n = x_1^n, X_2^n = x_2^n)$$
$$\geq \frac{1}{n} \sum_{(x_1^n, x_2^n) \in T_\epsilon^n[P_{X_1 X_2}]} Pr\{X_1^n = x_1^n, X_2^n = x_2^n\} H(Y_2^n | X_1^n = x_1^n, X_2^n = x_2^n)$$
$$= \frac{1}{n} \sum_{(x_1^n, x_2^n) \in T_\epsilon^n[P_{X_1 X_2}]} Pr\{X_1^n = x_1^n, X_2^n = x_2^n\}$$
$$\cdot \sum_{(a,b) \in \mathcal{X}_1 \times \mathcal{X}_2} N(a,b | x_1^n, x_2^n) \sum_{y_2 \in \mathcal{Y}_2} -p(y_2 | a,b) \log p(y_2 | a,b) \tag{129}$$
$$\overset{(a)}{\geq} \sum_{(x_1^n, x_2^n) \in T_\epsilon^n[P_{X_1 X_2}]} Pr\{X_1^n = x_1^n, X_2^n = x_2^n\}$$
$$\sum_{(a,b) \in \mathcal{X}_1 \times \mathcal{X}_2} \Big( P(X_1 = a, X_2 = b) - \epsilon \Big) \sum_{y_2 \in \mathcal{Y}_2} -p(y_2 | a,b) \log p(y_2 | a,b)$$
$$= \sum_{(x_1^n, x_2^n) \in T_\epsilon^n[P_{X_1 X_2}]} Pr\{X_1^n = x_1^n, X_2^n = x_2^n\} \Big( H(Y_2 | X_1, X_2) - O(\epsilon) \Big)$$
$$\overset{(b)}{\geq} (1-\epsilon) H(Y_2 | X_1, X_2) - O(\epsilon)$$
$$\geq H(Y_2 | X_1, X_2) - O(\epsilon)$$

where $O(\epsilon) \to 0$ as $\epsilon \to 0$. In the above equation, $(a)$ follows from the definition that $(x_1^n, x_2^n) \in T_\epsilon^n[P_{X_1X_2}]$. The inequality $(b)$ makes use of the following

$$\sum_{(x_1^n, x_2^n) \in T_\epsilon^n(P_{X_1X_2})} Pr\{X_1^n = x_1^n, X_2^n = x_2^n\} \geq 1 - p_e \geq 1 - \epsilon. \tag{130}$$

To compute the third term in (120), we define

$$\rho(w_0, w_1, y_2^n, x_2^n) = \begin{cases} x_{1,w_0ab}^n & \text{if there is a unique } b \text{ such that} \\ & (q_{w_0}^n, x_{1,w_0ab}^n, x_2^n, y_2^n) \in T_\epsilon^n(P_{QX_1X_2Y_2}) \\ \text{arbitrary} & \text{otherwise} \end{cases} \tag{131}$$

Then

$$\begin{aligned} Pr\{X_1^n &\neq \rho(W_0, W_1, Y_2^n, X_2^n)\} \\ &= \sum_{w_0, a, b, s, t} p_{w_0, a, b, s, t} Pr\{x_{1,w_0ab}^n \neq \rho(w_0, w_1, Y_2^n, x_{2,w_0st}^n) | w_0, a, b, s, t\} \\ &= \lambda_1 < \epsilon \end{aligned} \tag{132}$$

Therefore, by Fano's inequality, we obtain

$$\frac{1}{n} H(X_1^n | W_0, W_1, Y_2^n, X_2^n) \leq \frac{1}{n} \left(1 + \lambda_1 \log 2^{n(R_0 + R_1')}\right) < \epsilon_2 \tag{133}$$

where $\epsilon_2$ is small for sufficiently large $n$.

To compute the fourth term in (120), we define

$$\hat{y}_2^n = \begin{cases} y_2^n & \text{if } (q_{w_0}^n, x_2^n, y_2^n) \in T_\epsilon^n(P_{QX_2Y_2}) \\ \text{arbitrary} & \text{otherwise} \end{cases} \tag{134}$$

We then obtain

$$\begin{aligned} \frac{1}{n} H(Y_2^n | W_0, X_2^n) &= \frac{1}{n} \sum_{w_0, x_2^n} Pr\{W_0 = w_0, X_2^n = x_2^n\} H(Y_2^n | X_2^n = x_2^n, W_0 = w_0) \\ &\leq \frac{1}{n} \sum_{w_0, x_2^n} Pr\{W_0 = w_0, X_2^n = x_2^n\} H(Y_2^n, \hat{Y}_2^n | W_0 = w_0, X_2^n = x_2^n) \\ &= \frac{1}{n} \sum_{w_0, x_2^n} Pr\{W_0 = w_0, X_2^n = x_2^n\} \\ &\quad \cdot \left(H(\hat{Y}_2^n | W_0 = w_0, X_2^n = x_2^n) + H(Y_2^n | W_0 = w_0, X_2^n = x_2^n, \hat{Y}_2^n)\right) \end{aligned} \tag{135}$$

The first term in the preceding equation can be bounded as

$$\frac{1}{n}\sum_{w_0,x_2^n} Pr\{W_0 = w_0, X_2^n = x_2^n\} H(\hat{Y}_2^n | W_0 = w_0, X_2^n = x_2^n)$$
$$\leq \frac{1}{n} \sum_{w_0,x_2^n} Pr\{W_0 = w_0, X_2^n = x_2^n\} \log \|T_\epsilon^n(P_{Y_2|X_2,Q}|q_{w_0}^n, x_2^n)\|$$
$$\leq \sum_{w_0,x_2^n} Pr\{W_0 = w_0, X_2^n = x_2^n\}(H(Y_2|X_2,Q) + \epsilon)$$
$$\leq H(Y_2|X_2,Q) + \epsilon$$
(136)

To bound the second term in (135), we use Fano's inequality and obtain

$$\frac{1}{n}\sum_{w_0,x_2^n} Pr\{W_0 = w_0, X_2^n = x_2^n\} H(Y_2^n | W_0 = w_0, X_2^n = x_2^n, \hat{Y}_2^n)$$
$$\leq \frac{1}{n}\sum_{w_0,x_2^n} Pr\{W_0 = w_0, X_2^n = x_2^n\}\left(1 + Pr\{Y_2^n \neq \hat{Y}_2^n | W_0 = w_0, X_2^n = x_2^n\} \log |\mathcal{Y}_2|^n\right)$$
$$= \frac{1}{n} + \sum_{w_0,x_2^n} Pr\{W_0 = w_0, X_2^n = x_2^n\}$$
$$\cdot Pr\{(q_{w_0}^n, x_2^n, y_2^n) \notin T_\epsilon^n(P_{QX_2Y_2}) | W_0 = w_0, X_2^n = x_2^n\} \log |\mathcal{Y}_2|$$
$$\leq \frac{1}{n} + \sum_{w_0,x_1^n,x_2^n} Pr\{W_0 = w_0, X_2^n = x_2^n, X_1^n = x_1^n\}$$
$$\cdot Pr\{(q_{w_0}^n, x_1^n, x_2^n, y_2^n) \notin T_\epsilon^n(P_{QX_1X_2Y_2})\} \log |\mathcal{Y}_2|$$
$$\stackrel{(a)}{\leq} \frac{1}{n} + \sum_{\substack{w_0,x_1^n,x_2^n: \\ (q_{w_0}^n, x_1^n, x_2^n) \notin T_\epsilon^n}} Pr\{W_0 = w_0, X_2^n = x_2^n, X_1^n = x_1^n\} \log |\mathcal{Y}_2|$$
$$+ \sum_{\substack{w_0,x_1^n,x_2^n: \\ (q_{w_0}^n, x_1^n, x_2^n) \in T_\epsilon^n}} Pr\{W_0 = w_0, X_2^n = x_2^n, X_1^n = x_1^n\}$$
$$\cdot Pr\{(q_{w_0}^n, x_1^n, x_2^n, y_2^n) \notin T_\epsilon^n(P_{QX_1X_2Y_2})\} \log |\mathcal{Y}_2|$$
$$\leq \epsilon_3$$
(137)

where $\epsilon_3$ is small for sufficiently large $n$. In the preceding equation, the second term in $(a)$ is small because

$$\sum_{\substack{w_0,x_1^n,x_2^n: \\ (q_{w_0}^n, x_1^n, x_2^n) \notin T_\epsilon^n}} Pr\{W_0 = w_0, X_2^n = x_2^n, X_1^n = x_1^n\} \leq p_e \leq \epsilon.$$
(138)

Hence, the fourth term in (120) is

$$\frac{1}{n}H(Y_2^n|W_0, X_2^n) \leq H(Y_2|X_2, Q) + \epsilon_3 \tag{139}$$

Substituting (128), (129), (133) and (139) into (120), we obtain

$$\begin{aligned} H(W_1|Y_2^n, X_2^n, W_0, W_2) &\geq R_1' + H(Y_2|X_1, X_2) - H(Y_2|X_2, Q) - \epsilon_4 \\ &= R_1' + H(Y_2|X_1, X_2, Q) - H(Y_2|X_2, Q) - \epsilon_4 \\ &= R_1' - I(X_1; Y_2|X_2, Q) - \epsilon_4 \end{aligned} \tag{140}$$

where $\epsilon_4$ is small for sufficiently large $n$.

Similarly, we can also obtain

$$H(W_2|Y_1^n, X_1^n, W_0, W_1) \geq R_2' - I(X_2; Y_1|X_1, Q) - \epsilon_5 \tag{141}$$

where $\epsilon_5$ is small for sufficiently large $n$.

Hence, using the codebook given in Lemma 5 in Step 1 and the encoding functions defined in (116) in Step 2, the equivocation rates for sufficiently large $n$ are given by

$$R_{1,e} \leq R_1' - I(X_1; Y_2|X_2, Q) \tag{142}$$

and

$$R_{2,e} \leq R_2' - I(X_2; Y_1|X_1, Q). \tag{143}$$

**Step 4: Other Cases**

In Step 1, we have assumed that $R_1' > I(X_1; Y_2|X_2, Q)$ and $R_2' > I(X_2; Y_1|X_1, Q)$. If $R_1' \leq I(X_1; Y_2|X_2, Q)$, we generate $2^{nR_1}$ codewords $x_1^n$, and do not require $\lambda_1 < \epsilon$ in Lemma 5. We set $R_{1,e} = 0$. Similarly, if $R_2' \leq I(X_2; Y_1|X_1, Q)$, we set $R_{2,e} = 0$.

In Step 2, we have assumed that $R_1 > R_1' - I(X_1; Y_2|X_2, Q)$ and $R_2 > R_2' - I(X_2; Y_1|X_1, Q)$. If $R_1 \leq R_1' - I(X_1; Y_2|X_2, Q)$, we change the encoder $f_1$ to be the following:

$$\begin{aligned} f_1: \ & (W_0, W_1) \to \{x_{1,iab}^n\}, \text{ mapping } (w_0, w_1) \to x_{1,w_0w_1b}^n, \\ & \text{where } b \text{ is chosen uniformly from the set } [1, 2^{nI(X_1;Y_2|X_2,Q)}] \end{aligned} \tag{144}$$

In this case, note that the number of codewords is less than the number of rows in the codebook. The encoding strategy is to map each codeword to each row. It is expected that in this case the other user (user 2) is not able to decode any information, and hence user 1 achieves perfect secrecy.

In fact, the first term of the equivocation rate in (120) becomes

$$\frac{1}{n}H(X_1|W_0) = R_1 + I(X_1; Y_2|X_2, Q) \tag{145}$$

because for each $W_0 = i$, $X_1^n$ has $2^{n(R_1+I(X_1;Y_2|X_2,Q))}$ possible equally likely values. All other terms in (120) remain the same as before. We hence have

$$\begin{aligned} H(W_1|Y_2^n, X_2^n, W_0, W_2) \\ \geq R_1 + I(X_1;Y_2|X_2,Q) + H(Y_2|X_1,X_2) - H(Y_2|X_2,Q) - \epsilon \\ = R_1 + I(X_1;Y_2|X_2,Q) - I(X_1;Y_2|X_2,Q) - \epsilon \\ = R_1 - \epsilon. \end{aligned} \quad (146)$$

Thus, for sufficiently large $n$,

$$R_{1,e} \leq R_1, \quad (147)$$

and user 1 achieves perfect secrecy.

We can similarly obtain that if $R_2 \leq R_2' - I(X_2;Y_1|X_1,Q)$, user 2 achieves perfect secrecy, i.e.,

$$R_{2,e} \leq R_2. \quad (148)$$

Hence in summary, for a given point $(R_0, R_1, R_2) \in \tilde{\mathcal{C}}_{MAC}^p$, the achievable equivocation rate pairs are in the set $\tilde{\mathcal{S}}_e(R_0, R_1, R_2)$ defined in (90).

## 6.4 Proof of Theorem 8

In this section, we change $\mathcal{S}_e(R_0, R_1, R_2)$ given in (76) to an explicit form that is characterized by inequality bounds only. We need only to derive the explicit form for $\tilde{\mathcal{S}}_e(R_0, R_1, R_2)$ given in (90). The explicit form for $\mathcal{S}_e(R_0, R_1, R_2)$ follows by prefixing two discrete memoryless channels to the GMAC as reasoned at the beginning of Section 6.2.

We first characterize $\tilde{\mathcal{S}}_e(R_0, R_1, R_2)$ given in (90) in a more convenient form. We define the following set $\mathcal{B}(R_0, R_1, R_2)$ of $(R_1', R_2')$:

$$\mathcal{B}(R_0, R_1, R_2) := \left\{ \begin{array}{l} (R_1', R_2'): \\ R_1' \leq I(X_1;Y|X_2,Q), \\ R_1' + R_2 \leq I(X_1,X_2;Y|Q), \\ R_0 + R_1' + R_2 \leq I(X_1,X_2;Y) \\ R_2' \leq I(X_2;Y|X_1,Q), \\ R_1 + R_2' \leq I(X_1,X_2;Y|Q), \\ R_0 + R_1 + R_2' \leq I(X_1,X_2;Y) \\ R_1' + R_2' \leq I(X_1,X_2;Y|Q), \\ R_0 + R_1' + R_2' \leq I(X_1,X_2;Y) \end{array} \right\} \quad (149)$$

For $\tilde{\mathcal{S}}_e(R_0, R_1, R_2)$, if we replace union over the set $\mathcal{A}(R_0, R_1, R_2)$ with union over the set

$\mathcal{B}(R_0, R_1, R_2)$ as in the following, the region $\tilde{\mathcal{S}}_e(R_0, R_1, R_2)$ remains unchanged; i.e.,

$$\tilde{\mathcal{S}}_e(R_0, R_1, R_2) = \bigcup_{(R'_1, R'_2) \in \mathcal{B}(R_0, R_1, R_2)} \left\{ \begin{array}{l} (R_{1,e}, R_{2,e}) : \\ 0 \leq R_{1,e} \leq R_1 \\ R_{1,e} \leq [R'_1 - I(X_1; Y_2 | X_2, Q)]_+ \\ 0 \leq R_{2,e} \leq R_2 \\ R_{2,e} \leq [R'_2 - I(X_2; Y_1 | X_1, Q)]_+ \end{array} \right\}. \quad (150)$$

To see this, we plot the two sets $\mathcal{A}(R_0, R_1, R_2)$ and $\mathcal{B}(R_0, R_1, R_2)$ in Fig. 12. For any point $(r'_1, r'_2)$ that is in $\mathcal{B}(R_0, R_1, R_2)$ but not in $\mathcal{A}(R_0, R_1, R_2)$, there exists a corresponding point $(\bar{r}'_1, \bar{r}'_2) \in \mathcal{A}(R_0, R_1, R_2)$ such that $r'_1 \leq \bar{r}'_1$ and $r'_2 \leq \bar{r}'_2$. Hence

$$\left\{ \begin{array}{l} (R_{1,e}, R_{2,e}) : \\ 0 \leq R_{1,e} \leq R_1 \\ R_{1,e} \leq [r'_1 - I(X_1; Y_2 | X_2, Q)]_+ \\ 0 \leq R_{2,e} \leq R_2 \\ R_{2,e} \leq [r'_2 - I(X_2; Y_1 | X_1, Q)]_+ \end{array} \right\} \subset \left\{ \begin{array}{l} (R_{1,e}, R_{2,e}) : \\ 0 \leq R_{1,e} \leq R_1 \\ R_{1,e} \leq [\bar{r}'_1 - I(X_1; Y_2 | X_2, Q)]_+ \\ 0 \leq R_{2,e} \leq R_2 \\ R_{2,e} \leq [\bar{r}'_2 - I(X_2; Y_1 | X_1, Q)]_+ \end{array} \right\} \quad (151)$$

Therefore, those points that are in $\mathcal{B}(R_0, R_1, R_2)$ but not in $\mathcal{A}(R_0, R_1, R_2)$ do not contribute new $(R_{1,e}, R_{2,e})$ for $\tilde{\mathcal{S}}_e(R_0, R_1, R_2)$.

We now show that $\tilde{\mathcal{S}}_e(R_0, R_1, R_2)$ given in (150) is equivalent to the following region $\tilde{\mathcal{S}}'_e(R_0, R_1, R_2)$.

$$\tilde{\mathcal{S}}'_e = \tilde{\mathcal{L}}_1 \cup \tilde{\mathcal{L}}_2 \cup \tilde{\mathcal{L}}_3 \quad (152)$$

where

$$\tilde{\mathcal{L}}_1 = \left\{ \begin{array}{l} (R_{1,e}, R_{2,e}) : \\ 0 \leq R_{1,e} \leq R_1, \ 0 \leq R_{2,e} \leq R_2 \\ R_{1,e} \leq [I(X_1; Y | X_2, Q) - I(X_1; Y_2 | X_2, Q)]_+, \\ R_{1,e} \leq [I(X_1, X_2; Y | Q) - R_2 - I(X_1; Y_2 | X_2, Q)]_+, \\ R_{1,e} \leq [I(X_1, X_2; Y) - R_0 - R_2 - I(X_1; Y_2 | X_2, Q)]_+ \\ R_{2,e} \leq [I(X_2; Y | X_1, Q) - I(X_2; Y_1 | X_1, Q)]_+, \\ R_{2,e} \leq [I(X_1, X_2; Y | Q) - R_1 - I(X_2; Y_1 | X_1, Q)]_+, \\ R_{2,e} \leq [I(X_1, X_2; Y) - R_0 - R_1 - I(X_2; Y_1 | X_1, Q)]_+ \\ R_{1,e} + R_{2,e} \leq [I(X_1, X_2; Y | Q) - I(X_1; Y_2 | X_2, Q) - I(X_2; Y_1 | X_1, Q)]_+, \\ R_{1,e} + R_{2,e} \leq [I(X_1, X_2; Y) - R_0 - I(X_1; Y_2 | X_2, Q) - I(X_2; Y_1 | X_1, Q)]_+ \end{array} \right\} \quad (153)$$

$$\tilde{\mathcal{L}}_2 = \left\{ \begin{array}{l} (R_{1,e}, R_{2,e}) : \\ 0 \leq R_{1,e} \leq R_1, \\ R_{1,e} \leq [I(X_1; Y | X_2, Q) - I(X_1; Y_2 | X_2, Q)]_+, \\ R_{1,e} \leq [I(X_1, X_2; Y | Q) - R_2 - I(X_1; Y_2 | X_2, Q)]_+, \\ R_{1,e} \leq [I(X_1, X_2; Y) - R_0 - R_2 - I(X_1; Y_2 | X_2, Q)]_+ \\ R_{2,e} = 0 \end{array} \right\} \quad (154)$$

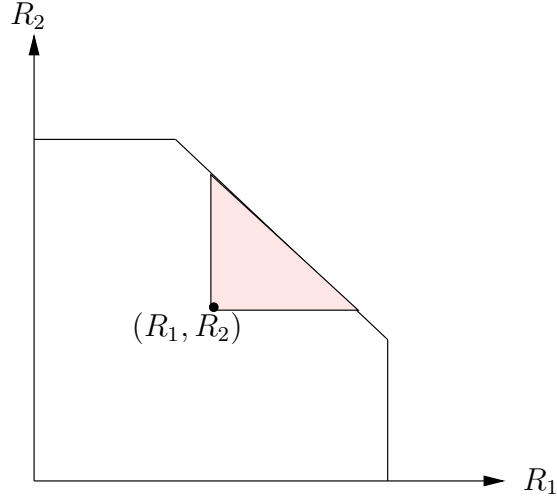

(a): Region $\mathcal{A}(R_0, R_1, R_2)$ of $(R'_1, R'_2)$.

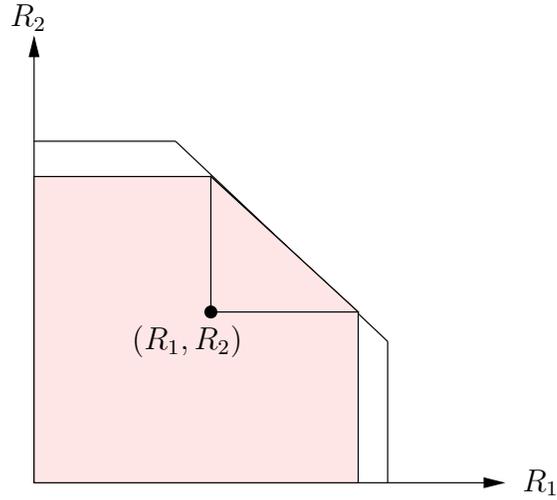

(b): Region $\mathcal{B}(R_0, R_1, R_2)$ of $(R'_1, R'_2)$

Figure 12: Regions $\mathcal{A}(R_0, R_1, R_2)$ and $\mathcal{B}(R_0, R_1, R_2)$ of $(R'_1, R'_2)$

and
$$\tilde{\mathcal{L}}_3 = \left\{ \begin{array}{l} (R_{1,e}, R_{2,e}) : \\ R_{1,e} = 0, \\ 0 \leq R_{2,e} \leq R_2, \\ R_{2,e} \leq [I(X_2; Y|X_1, Q) - I(X_2; Y_1|X_1, Q)]_+, \\ R_{2,e} \leq [I(X_1, X_2; Y|Q) - R_1 - I(X_2; Y_1|X_1, Q)]_+, \\ R_{2,e} \leq [I(X_1, X_2; Y) - R_0 - R_1 - I(X_2; Y_1|X_1, Q)]_+ \end{array} \right\} \qquad (155)$$

We first show $\tilde{\mathcal{S}}_e(R_0, R_1, R_2) \subset \tilde{\mathcal{S}}'_e(R_0, R_1, R_2)$. For a point $(r_{1,e}, r_{2,e}) \in \tilde{\mathcal{S}}_e(R_0, R_1, R_2)$, we consider the following four cases.

Case 1. If $r_{1,e} > 0$ and $r_{2,e} > 0$, then there exits $(R_1', R_2') \in \mathcal{B}(R_0, R_1, R_2)$, such that

$$0 < r_{1,e} \leq R_1' - I(X_1; Y_2 | X_2, Q) \tag{156}$$

and

$$0 < r_{2,e} \leq R_2' - I(X_2; Y_1 | X_1, Q). \tag{157}$$

Applying the bounds that $(R_1', R_2') \in \mathcal{B}(R_0, R_1, R_2)$ needs to satisfy to the preceding inequalities, it is clear that $(r_{1,e}, r_{2,e}) \in \tilde{\mathcal{L}}_1$.

Case 2. If $r_{1,e} > 0$ and $r_{2,e} = 0$, then it is easy to check that $(r_{1,e}, r_{2,e}) \in \tilde{\mathcal{L}}_2$.

Case 3. If $r_{1,e} = 0$ and $r_{2,e} > 0$, then it is easy to check that $(r_{1,e}, r_{2,e}) \in \tilde{\mathcal{L}}_3$.

Case 4. If $r_{1,e} = 0$ and $r_{2,e} = 0$, then it is trivially true that $(r_{1,e}, r_{2,e}) \in \tilde{\mathcal{S}}_e'(R_0, R_1, R_2)$.

We now show that $\tilde{\mathcal{S}}_e'(R_0, R_1, R_2) \subset \tilde{\mathcal{S}}_e(R_0, R_1, R_2)$. We first show $\tilde{\mathcal{L}}_1(R_0, R_1, R_2) \subset \tilde{\mathcal{S}}_e(R_0, R_1, R_2)$. We consider the following four cases.

Case 1. If $r_{1,e} > 0$ and $r_{2,e} = 0$, then we let $r_1' = r_{1,e} + I(X_1; Y_2 | X_2, Q)$ and $r_2' = R_2$. From the first three bounds that define $\tilde{\mathcal{L}}_1(R_0, R_1, R_2)$ and $(R_0, R_1, R_2) \in \tilde{\mathcal{C}}_{MAC}^p$, it is easy to check that $(r_1', r_2') \in \mathcal{B}(R_0, R_1, R_2)$. Hence $(r_{1,e}, r_{2,e}) \in \tilde{\mathcal{S}}_e(R_0, R_1, R_2)$.

Case 2. If $r_{1,e} = 0$ and $r_{2,e} > 0$, then it can be similarly checked that $(r_{1,e}, r_{2,e}) \in \tilde{\mathcal{S}}_e(R_0, R_1, R_2)$ as in Case 1.

Case 3. If $r_{1,e} > 0$ and $r_{2,e} > 0$, then let $r_1' = r_{1,e} + I(X_1; Y_2 | X_2, Q)$ and $r_2' = r_{2,e} + I(X_2; Y_2 | X_1, Q)$. It is easy to check that $(r_1', r_2') \in \mathcal{B}(R_0, R_1, R_2)$. Hence $(r_{1,e}, r_{2,e}) \in \tilde{\mathcal{S}}_e(R_0, R_1, R_2)$.

Case 4. If $r_{1,e} = 0$ and $r_{2,e} = 0$, then it is trivially true that $(r_{1,e}, r_{2,e}) \in \tilde{\mathcal{S}}_e(R_0, R_1, R_2)$.

Finally, the conditions $\tilde{\mathcal{L}}_2(R_0, R_1, R_2) \subset \tilde{\mathcal{S}}_e(R_0, R_1, R_2)$ and $\tilde{\mathcal{L}}_2(R_0, R_1, R_2) \subset \tilde{\mathcal{S}}_e(R_0, R_1, R_2)$ can be similarly shown as in Case 1 and Case 2, respectively.

# 7 Conclusions

We have studied the capacity-equivocation region of the GMAC with confidential messages. For the GMAC with one confidential message set, we have derived inner and outer bounds on the capacity-equivocation region. Although the two bounds only match partially, they are tight enough to characterize the secrecy capacity region, where confidential messages sent by user 1 are perfectly hidden from user 2. For the degraded GMAC, we have established the capacity-equivocation region. We have further derived the capacity-equivocation region for two examples of degraded GMACs. In particular, we have found that the capacity-equivocation region of GMACs with confidential messages depends only on the marginal channels $p(y|x_1, x_2)$, $p(y_1|x_1, x_2)$, and $p(y_2|x_1, x_2)$. Based on this observation, we have obtained the capacity-equivocation region for the Gaussian GMAC (not necessarily physically degraded) with one confidential message set.

We have also obtained an achievable rate-equivocation region (inner bound on the capacity-equivocation region) for the general case of the GMAC with two confidential message sets. The region takes a much more complicated form than the case of the GMAC with one confidential message set. We have further derived an equivalent but explicit form for the achievable rate-equivocation region. Moreover, we have shown that the achievable rate-equivocation region for the case of two confidential message sets carries a new feature of a trade-off between the two equivocation rates corresponding to the two confidential message sets sent by the two users.